\documentclass[prd,12pt]{article}

\usepackage{tikz,xcolor,lscape}

\usepackage{amsmath,amssymb,graphicx,multirow,xspace,slashed}
\usepackage[colorlinks=true,linktocpage=true,linkcolor=purple,citecolor=UOgreen,urlcolor=purple]{hyperref} 
\usepackage[compress,numbers]{natbib}
\usepackage{subfigure, placeins}
\usepackage{booktabs}
\usepackage{blindtext, rotating}
\usepackage{afterpage}
\usepackage{enumitem}
\usepackage{ marvosym }
\usepackage{authblk} 
\usepackage{verbatim}
\usepackage{soul} 
\usepackage[normalem]{ulem}
\usepackage{pifont}
\usepackage[usestackEOL]{stackengine}
\usepackage{subfiles}
\usepackage{fontawesome}
\newcommand\sq{\framebox(10,10){}\kern\fboxrule}

\setstackgap{S}{\fboxrule}

\usepackage[utf8]{inputenc}

\usepackage{floatrow}
\newfloatcommand{capbtabbox}{table}[][\FBwidth]

\usepackage[font=footnotesize,labelfont=bf]{caption}

\usepackage{lineno}

\usepackage{ulem}

\allowdisplaybreaks

\graphicspath{{figs/}}
\usepackage{setspace}
\usepackage{adjustbox}

\long\def\symbolfootnote[#1]#2{\begingroup%
\def\thefootnote{\fnsymbol{footnote}}\footnote[#1]{#2}\endgroup}

\makeatletter
\newcommand{\github}[1]{%
   \href{#1}{\faGithubSquare}%
}
\makeatother

\newcommand{\newc}{\newcommand}
\newc{\gsim}{\lower.7ex\hbox{$\;\stackrel{\textstyle>}{\sim}\;$}}
\newc{\lsim}{\lower.7ex\hbox{$\;\stackrel{\textstyle<}{\sim}\;$}}
\newc{\gev}{\,{\rm GeV}}
\newc{\mev}{\,{\rm MeV}}
\newc{\ev}{\,{\rm eV}}
\newc{\kev}{\,{\rm keV}}
\newc{\tev}{\,{\rm TeV}}
\newc{\MHT}{$H_T^{\text{miss}}$}
\newc{\MET}{$\slashed{E}_T$}
\newc{\MTT}{$M_{T2}$}
\newcommand{\sv}{\langle \sigma v \rangle}

\newcommand{\ypfo}{Y_{\phi}^{\textrm{FO}}}
\newcommand{\ycfi}{Y_{\chi}^{\textrm{FI}}}

\newc{\mz}{M_Z}
\newc{\mpl}{M_*}
\newc{\mw}{m_{\rm weak}}
\newc{\nr}[1]{N^c_R{}_{#1}}

\def\beq{\begin{equation}}
\def\eeq{\end{equation}}
\newcommand{\bea}{\begin{eqnarray}\begin{aligned}}
\newcommand{\eea}{\end{aligned}\end{eqnarray}}
\def\bitem{\begin{itemize}}
\def\eitem{\end{itemize}}

\definecolor{UOgreen}{cmyk}{.96,.26,1.00,.15}
\definecolor{darkgreen}{rgb}{0,0.5,0}
\definecolor{goodyellow}{rgb}{0.9,0.7,0}
\definecolor{DarkGreen}{rgb}{0,0.6,0}
\definecolor{goodgreen}{rgb}{0,.6,0.4}

\numberwithin{equation}{section}

\newcommand\fverb{\setbox\fverbbox=\hbox\bgroup\verb}

\newbox\fverbbox

\begin{document}
\baselineskip 0.6cm

\begin{titlepage}

\thispagestyle{empty}

\begin{center}

\vskip 0.1cm

\hspace{4.8in}

\vspace{1.cm}

{\Large \bf A Duet of Freeze-in and Freeze-out: \\ Lepton-Flavored Dark Matter and Muon Colliders}

\vskip 0.5cm

\vskip 0.5cm

\vskip 1.0cm
{\large Pouya Asadi, Aria Radick, Tien-Tien Yu}
\vskip 1.0cm
\textit{Institute for Fundamental Science and Department of Physics, \\ University of Oregon, Eugene, OR 97403, USA} 
\vskip 0.3cm

\end{center}

\vskip 0.6cm

\begin{abstract}

We study a Lepton-Flavored Dark Matter model and its signatures at a future Muon Collider. 
We focus on the less-explored regime of feeble dark matter interactions, which suppresses the dangerous lepton-flavor violating processes, gives rise to dark matter freeze-in production, and leads to long-lived particle signatures at colliders.
We find that the interplay of dark matter freeze-in and its mediator freeze-out gives rise to an upper bound of around TeV scales on the dark matter mass. The signatures of this model depend on the lifetime of the mediator, and can range from generic prompt decays to more exotic long-lived particle signals. 
In the prompt region, we calculate the signal yield, study useful kinematics cuts, and report tolerable systematics that would allow for a $5\sigma$ discovery. 
In the long-lived region, we calculate the number of charged tracks and displaced lepton signals of our model in different parts of the detector, and uncover kinematic features that can be used for background rejection. 
We show that, unlike in hadron colliders, multiple production channels contribute significantly which leads to sharply distinct kinematics for electroweakly-charged long-lived particle signals. 
Ultimately, the collider signatures of this lepton-flavored dark matter model are common amongst models of electroweak-charged new physics, rendering this model a useful and broadly applicable benchmark model for future Muon Collider studies that can help inform work on detector design and studies of systematics. \github{https://github.com/ariaradick/LFDM_at_MuC}

\end{abstract}

\flushbottom

\end{titlepage}

\setcounter{page}{1}

\tableofcontents

\vskip 1cm

\newpage

\begin{spacing}{1.25}

\section{Introduction}
\label{sec:intro}

The Standard Model (SM) of particle physics is an extremely successful model, making precise and accurate predictions for a wide range of phenomena. 
However, it fails to explain several fundamental phenomena. This includes explanations for the origin of the electroweak scale, the existence of three flavors of matter and their mass hierarchies, and the particle nature of Dark Matter (DM), to name a few. 

Many studies are undertaken in pursuit of an answer to these questions. 
On the experimental front, collider experiments are unique in their ability in probing a multitude of different models and unveiling various aspects of potential beyond SM (BSM) particles above the electroweak scale.

To that end, various future collider experiments are proposed to push the energy frontier. One such option that, thanks to increasing community interest (see {\it e.g.}~\cite{MuonCollider:2022nsa}), has risen as a compelling contender is a Muon Collider (MuC) \cite{Ankenbrandt:1999cta,Wang_2016,Boscolo:2018ytm,Neuffer:2018yof,Delahaye:2019omf,Shiltsev:2019rfl}. 
There are considerable technical challenges that need to be overcome before a viable high energy MuC can be constructed, see Refs.~\cite{MuonCollider:2022xlm,MuonCollider:2022glg,MuonCollider:2022nsa,MuonCollider:2022ded,Black:2022cth,Accettura:2023ked} for recent reviews. 
Nonetheless, many theoretical studies are carried out to underscore the reach of such a machine in probing different BSM models and a  template of well-motivated targets are being developed for searches at such a facility.
The clean environment of a lepton collider, combined with substantial parton distribution function (PDF) of electroweak gauge bosons in a muon at TeV scales, turns such a machine into a suitable candidate for an in-depth study of the Higgs and precision electroweak measurements \cite{Eichten:2013ckl,Chakrabarty:2014pja,Buttazzo:2018qqp,Chiesa:2020awd,Han:2020pif,Bandyopadhyay:2020otm,Buttazzo:2020uzc,Liu:2021jyc,Han:2021udl,Franceschini:2021aqd,Chiesa:2021qpr,Chen:2022msz,Spor:2022hhn,Amarkhail:2023xsc}, flavorful new physics \cite{Capdevilla:2020qel,Buttazzo:2020ibd,Yin:2020afe,Huang:2021nkl,Capdevilla:2021rwo,Chen:2021rnl,Huang:2021biu,Asadi:2021gah,Bandyopadhyay:2021pld,Qian:2021ihf,Homiller:2022iax,Azatov:2022itm,Yang:2023ojm,Altmannshofer:2023uci,Jana:2023ogd,Ghosh:2023xbj}, DM candidates \cite{Han:2020uak,Bottaro:2021srh,Liu:2022byu,Jueid:2023zxx,Vignaroli:2023rxr,Belfkir:2023vpo}, as well as other general SM and BSM physics studies \cite{DiLuzio:2018jwd,Costantini:2020stv,Liu:2021akf,AlAli:2021let,Casarsa:2021rud,Bao:2022onq,Inan:2022rcr,Lv:2022pts,Chen:2022yiu,Kwok:2023dck,Inan:2023pva,Chowdhury:2023imd,Belfkir:2023lot,Chigusa:2023rrz}. There are also 
proposals for earlier stage facilities, such as a beam dump with accelerated muon beams, and interesting new physics that they can probe \cite{Cesarotti:2022ttv,Cesarotti:2023sje}.

At the moment there exists no final detector design for such a future machine. 
Theoretical works tabulating well-motivated models can inform development of such a design. 
In particular, there are many well-motivated DM models that can be searched for at a MuC, including various minimal Weakly-Interacting Massive Particle (WIMP) DM models \cite{Han:2020uak}. 
Many of these DM models, including WIMPs, are strongly constrained by direct detection searches.

In this work, we focus on a less-constrained -- yet similarly minimal -- class of DM models, namely flavored DM~\cite{Agrawal:2011ze,Batell:2013zwa,Agrawal:2014una,Agrawal:2014aoa,Agrawal:2015tfa,Agrawal:2015kje,Agrawal:2016uwf,Desai:2020rwz,Acaroglu:2022hrm}. 
In such setups, the DM, which is a singlet fermion of the SM, interacts with the SM only via a new scalar mediator with the same charges as one of the SM fermion fields. 
We focus on the setup where the new mediator has the same charges as right-handed (RH) leptons of the SM, and study its relic abundance calculation and signals at a MuC. 

Previous works on flavored DM models have focused on DM production through thermal freeze-out~\cite{Agrawal:2011ze,Batell:2013zwa,Agrawal:2014una,Agrawal:2014aoa,Agrawal:2015tfa,Agrawal:2015kje,Agrawal:2016uwf,Desai:2020rwz,Acaroglu:2022hrm}. 
A sub-TeV DM mass gives rise to the correct relic abundance in such setups. 
DM coupling to SM leptons should be aligned with SM lepton Yukawa matrix in order to suppress dangerous flavor-changing neutral current (FCNC) and lepton-flavor violation (LFV) processes. 
These constraints can be abated by decreasing the size of the DM-mediator Yukawa coupling, which opens up a new mechanism for DM production: freeze-in~\cite{Hall:2009bx}. 
The smallness of the Yukawa coupling also leads to a rich set of collider signatures, ranging from prompt to long-lived particle (LLP). 

In this work, we will study in detail the production mechanism for freeze-in flavored DM, demonstrating how the interplay between mediator and DM abundance leads to a bounded parameter space, as well as the collider signatures at a future MuC. 
In particular, given the generic nature of the production rate of our model, our signal yield can serve as a benchmark target that will inform future experimental studies.

In Section~\ref{sec:model} we introduce the flavored DM model that we study in this work. We go through its relic abundance calculation in the freeze-in regime, as well as a review of some existing LHC bounds on its parameter space. 
We will then comment on its production channels at a future MuC in Section~\ref{sec:main}. 
We will then study the kinematics of this model and prospects for its discovery in parts of its parameter space where it decays promptly in the detector in Section~\ref{sec:prompt}. 
We will also study its kinematics and signal yield in the LLP part of the parameter space in Section~\ref{sec:displaced}, before concluding in Section~\ref{sec:conclusion}. 
We also provide details about relic abundance calculation in Appendix~\ref{app:abundance}. 
Further histograms and explanations about the prompt (LLP) signal are included in Appendix~\ref{app:prompt} (\ref{app:LLP})

\section{Lepton-Flavored Dark Matter Model}
\label{sec:model}

We start by introducing the model under study in this paper. 
We augment the SM with a scalar mediator, $\phi$, that has the same charges under SM gauge groups as RH charged leptons, and neutral fermion $\chi$,\footnote{With the same field content we could also have $\phi L L$ and $H \bar{\chi} L$ terms that could potentially destabilize DM. We postulate a $\mathbb{Z}_2$ parity acting on $\phi$, $\chi$, and $\bar{\chi}$ that forbids these terms.}
\begin{equation}
\mathcal{L} \supset - m_\chi \bar{\chi}_\alpha \chi^\alpha - m_\phi^2 |\phi|^2 - \lambda_{i,\alpha} \phi \bar{e}_i \bar{\chi}_{\alpha}.
    \label{eq:L}
\end{equation}
Here $i$ ($\alpha$) is SM fermion (DM) flavor index, $\bar{e}$ denotes SM charged leptons of RH chirality, $\chi$ is the fermionic DM candidate, $\phi$ is the mediator that has the same charges under SM gauge group as a RH charged lepton, and $m_\chi$ ($m_\phi$) denotes the shared DM (mediator) mass. 
While having multiple DM flavors can give rise to interesting physics \cite{Agrawal:2011ze,Batell:2013zwa,Agrawal:2014una,Agrawal:2014aoa,Agrawal:2015tfa,Agrawal:2015kje,Agrawal:2016uwf,Desai:2020rwz}, we will consider only one DM flavor as multiple flavors will not affect our collider study.

It should be noted that such setups are studied under other names such as Effective WIMP \cite{Chang:2013oia,Chang:2014tea}, Fermion Portal DM \cite{Bai:2013iqa,Bai:2014osa}, and colored mediator models \cite{DiFranzo:2013vra,Garny:2014waa} as well. 
Flavored DM model can be considered a more general framework compared to these models since it also includes multiple DM flavor indices. 
Thus, we use the umbrella term flavored DM for our model, even though having multiple DM flavors, for the most part, does not change our study. 
We should also note that we focus on the case where the neutral particle, $\chi$, is a fermion. We could also consider the scenario in which this DM candidate is scalar and its mediator to SM is a charged heavier fermion.

In the existing studies of such setups (including a recent study of its signal at a MuC \cite{Jueid:2023zxx}), DM is assumed to be a thermal relic, freezing out of the SM bath at some point in the early universe. 
Nonetheless, DM abundance could also be determined via the freeze-in mechanism \cite{Hall:2009bx} instead,\footnote{The original freeze-in proposal is now known as IR freeze-in; see Ref.~\cite{Elahi:2014fsa} for what is known as UV freeze-in.} where DM never reaches thermal equilibrium with SM. 
While there exists some studies of flavored DM abundance via UV freeze-in \cite{DAmbrosio:2021wpd} or LHC signatures in this region of the parameter space \cite{Evans:2016zau,Calibbi:2021fld}, a proper study of the interplay between the DM freeze-in and its mediator's freeze-out, and their implications for collider searches are still undelivered.

In the upcoming sections, we will study the relic abundance of DM in such a setup more carefully, highlighting an interesting interplay of freeze-in and freeze-out which gives rise to an upper bound on DM mass and interesting signals at future colliders. 
It should be noted that we are implicitly assuming the reheating temperature in the early universe is above the mediator mass $m_\phi$, such that there initially exists an abundance of $\phi$ in equilibrium with SM, which subsequently freezes out before decaying to DM $\chi$. 
If the reheat temperature is below $\phi$ mass, DM is only produced via UV freeze-in \cite{Elahi:2014fsa} and it will not have an upper bound on its mass.

The original flavored DM setups required an artificial introduction of a non-trivial flavor ansatz to suppress dangerous FCNC and LFV. In the freeze-in regime, smallness of the $\lambda$ Yukawa coupling guarantees an automatic suppression of these processes. 
We will leave explorations of a natural realization of such small Yukawa couplings in the UV to future works and focus on their phenomenology at colliders in this work.

\subsection{Relic Abundance}
\label{subsec:abundance}

In this section we go through the relic abundance calculation of our setup. 
We will show that the interplay between the freeze-in and the freeze-out processes bounds our viable parameter space from all directions. 
Further details about our calculation of the relic abundance can be found in Appendix~\ref{app:abundance}; see also Refs.~\cite{Garny:2018ali,Decant:2021mhj} for a similar relic abundance calculation in a quark-flavored DM model.

\begin{figure}
    \centering
    \resizebox{0.99\columnwidth}{!}{
    \includegraphics{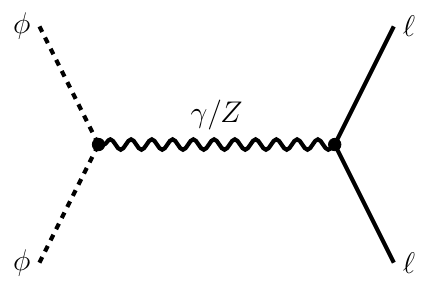}
    \hspace{0.2cm}
    \includegraphics{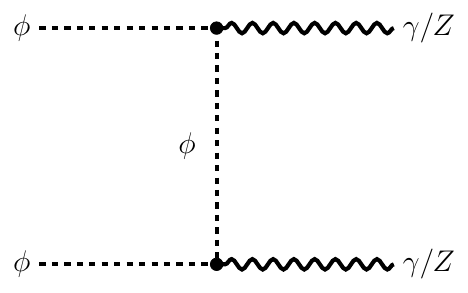}
    \hspace{0.2cm}
    \includegraphics{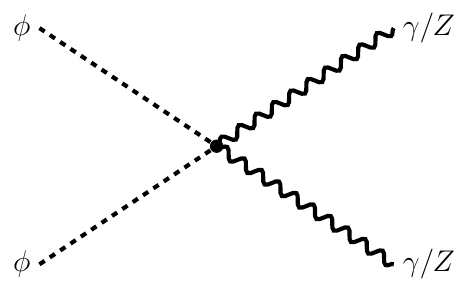}
    }\\
    \resizebox{0.28\columnwidth}{!}{
    \includegraphics{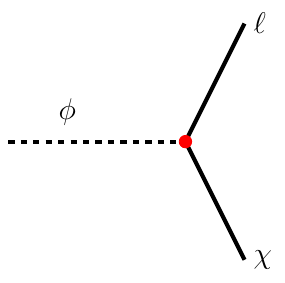}
    }
    \caption{Some diagrams relevant for Boltzmann equations governing freeze-out of $\phi$ (\textbf{top}) and the freeze-in of $\chi$ (\textbf{bottom}). The freeze-out diagrams are controlled by the electroweak gauge coupling (the black dots), while the freeze-in process is proportional to the small Yukawa coupling $\lambda$ (red dot). We assume comparable coupling to different lepton flavors in our calculation.  Other diagrams with higher number of $\lambda$ vertices or particle multiplicity are suppressed.}
    \label{fig:relic_diagrams}
\end{figure}

We start by reasserting that $\phi$ reaches equilibrium in the early universe, {\it i.e.} the reheat temperature is higher than the $\phi$ mass.
As a result, we should solve the coupled set of Boltzmann equations governing the abundances. 
Neglecting Pauli blocking and Bose enhancement, and assuming symmetric particle--anti-particle abundances, we can rewrite the evolution equations for $\phi$ and $\chi$ abundances, as~\cite{Kolb:1990vq,Hall:2009bx} 
\begin{align}
  \frac{d Y_\phi}{d x} = &- \sum_{\cal F} \frac{m_\phi^3 \sv_{\phi \phi \to {\cal F}}}{H(m_\phi) x^2} \left[ Y_\phi^2 - Y_{\phi, \textrm{EQ}}^2 \right] \nonumber \\
  &- \sum_\ell \frac{x^3}{H(m_\phi)} \frac{g_\phi \Gamma_{\phi \to \ell \chi}}{2 \pi^2} K_1(x) \left[ \frac{Y_\phi}{Y_{\phi, \textrm{EQ}}} - \frac{Y_\chi}{Y_{\chi, \textrm{EQ}}} \right], \\
  \frac{d Y_\chi}{d x} = & \sum_\ell \frac{x^3}{H(m_\phi)} \frac{g_\phi \Gamma_{\phi \to \ell \chi}}{2 \pi^2} K_1(x) \left[ \frac{Y_\phi}{Y_{\phi, \textrm{EQ}}} - \frac{Y_\chi}{Y_{\chi, \textrm{EQ}}} \right],
\end{align}
where $H(m_\phi)=1.66\sqrt{g_*^\rho}m_\phi^2/M_{\rm Pl}$ is the Hubble constant at $m_\phi$ and $x = m_\phi / T$. $Y_X = n_X / T^3$ is the yield of particle $X$ with $n_X$ the number density for particle $X$, and $Y_{X, \mathrm{EQ}}$ is the equilibrium value of $Y_X$. The expression for $Y_\phi$ consists of two parts that contribute to the depletion of $\phi$: the first line corresponds to the annihilation of $\phi$, where $\sv_{\phi \phi\to {\cal F}}$ is the thermally averaged cross-section for $\phi^+ \phi^-\to {\cal F}$ with ${\cal F} \in [\ell\bar \ell, q\bar q, \gamma \gamma, ZZ]$, while the second line corresponds to the decay of $\phi$, where $\Gamma_{\phi \to \ell \chi}$ is the decay width, $g_\phi=1$ is the number of spin degrees of freedom of $\phi$, and $K_1$ is the first order modified Bessel function of the second kind. The expression for $Y_\chi$ has one term which corresponds to the production of $\chi$ from $\phi\to\ell\chi$. In the calculations that follow, we assume the couplings $\lambda_{i} = \lambda$ are all equal. 
In Figure \ref{fig:relic_diagrams} we see the relevant diagrams for our calculation. 
In this freeze-in calculation other diagrams involving more $\lambda$ couplings or additional external particles are suppressed and negligible \cite{Hall:2009bx}.

\begin{figure}
    \centering
    \resizebox{0.7\columnwidth}{!}{
    \includegraphics{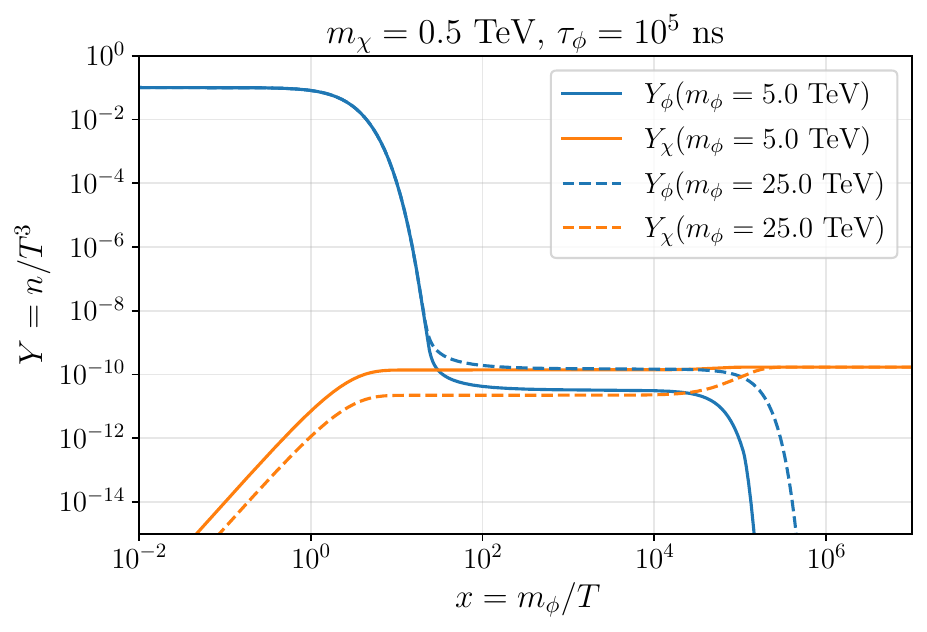}
    }
    \caption{ Abundance of $\chi$ (orange) and $\phi$ (blue) for the two $m_\phi$ values that result in today DM abundance for $\chi$ with the given $m_\chi$ and $\phi$ lifetime. Solid (dashed) lines correspond to the light (heavy) $\phi$ scenarios. The figure shows that we can have scenarios where $\chi$ abundance today is mostly dominated by freeze-in of $\chi$ itself (solid) or freeze-out, and subsequent decay, of the mediator $\phi$ (dashed). It also shows that with fixed lifetime and DM mass, there are two different mediator masses (and $\lambda$ couplings) that give rise to the right relic abundance. }
    \label{fig:boltz_solve}
\end{figure}

We numerically solve these coupled differential equations. 
In Figure \ref{fig:boltz_solve} we see the results for benchmark masses and a fixed mediator lifetime. 
We have chosen the Yukawa coupling $\lambda$ such that we get the correct final $\chi$ relic abundance. 
From this figure, we see that there are two scenarios that can give rise to the same final $\chi$ abundance. 
For the lighter $\phi$, freeze-in of $\chi$ dominates its final abundance, gaining merely a small boost when $\phi$ decays, while for a heavy $\phi$ we see that freeze-out of $\phi$ and its subsequent decay to $\chi$ dominates the final $\chi$ abundance.

We can then use the final $\chi$ yield to calculate its expected relic abundance. Assuming that $\chi$ constitutes all of the DM we set $\Omega_\chi h^2 = 0.12$, \textit{e.g.} see Ref.~\cite{Planck:2018vyg}, and find the value of the coupling $\lambda = \lambda'$ that recovers this for each pair of $\phi$ and $\chi$ masses. 
In our analysis we assume the same coupling between DM and all SM RH leptons. 
In Appendix \ref{app:abundance} we detail our semi-analytic approximation for calculating $\lambda'$.

\begin{figure}
    \centering
    \resizebox{0.8\columnwidth}{!}{
    \includegraphics{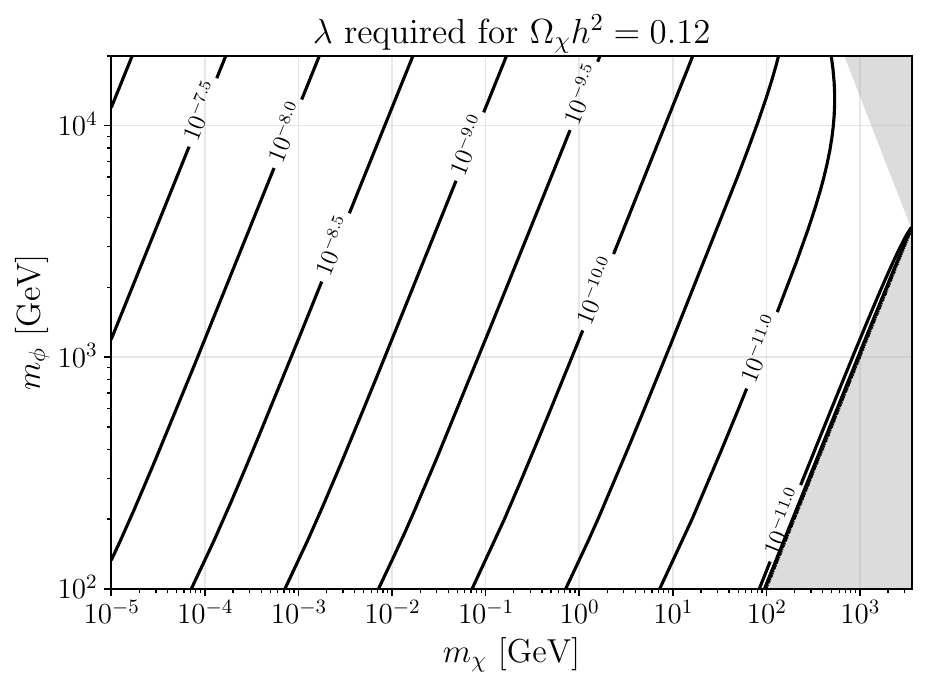}
    }
    \caption{The DM Yukawa coupling, $\lambda'$, that gives rise to the right relic abundance today. We assume the coupling to each fermion is the same. }
    \label{fig:lambda_parameter}
\end{figure}

In Figure \ref{fig:lambda_parameter} we show contours of $\lambda'$ for every point on the mass plane.
For large enough $m_\phi$ values the $\phi$ freeze-out cross-section will be small enough that we are left with too many $\phi$ particles after their freeze-out such that, after their decay to $\chi$, the universe will be overclosed. 
As we decrease $m_\phi$, and for a fixed $m_\chi$, at some point the freeze-out $\phi$ abundance will be low enough that, after decaying to $\chi$, we find the correct relic abundance today. 
The contribution from DM freeze-in to the abundance today will be sub-dominant. 
For lower $m_\phi$ values, the DM abundance from $\phi$ freeze-out and decay is augmented by a non-negligible freeze-in contribution to get the right relic abundance today.  
Stability of DM also puts a lower bound on the mediator mass $m_\phi \geqslant m_\chi$.

The upper bound on the mediator mass (from DM relic abundance) and the lower bound on it (from DM stability), suggests a bounded viable parameter space. 
This also implies an upper bound on the DM mass. 
Our relic abundance calculation shows that this upper bound is 
\begin{equation}
m_\chi \lesssim 3.6~\mathrm{TeV}.
    \label{eq:DMupper}
\end{equation}

\subsection{Existing Constraints}
\label{sec:lhcbounds}

\begin{figure}
    \centering
    \resizebox{0.8\columnwidth}{!}{
    \includegraphics{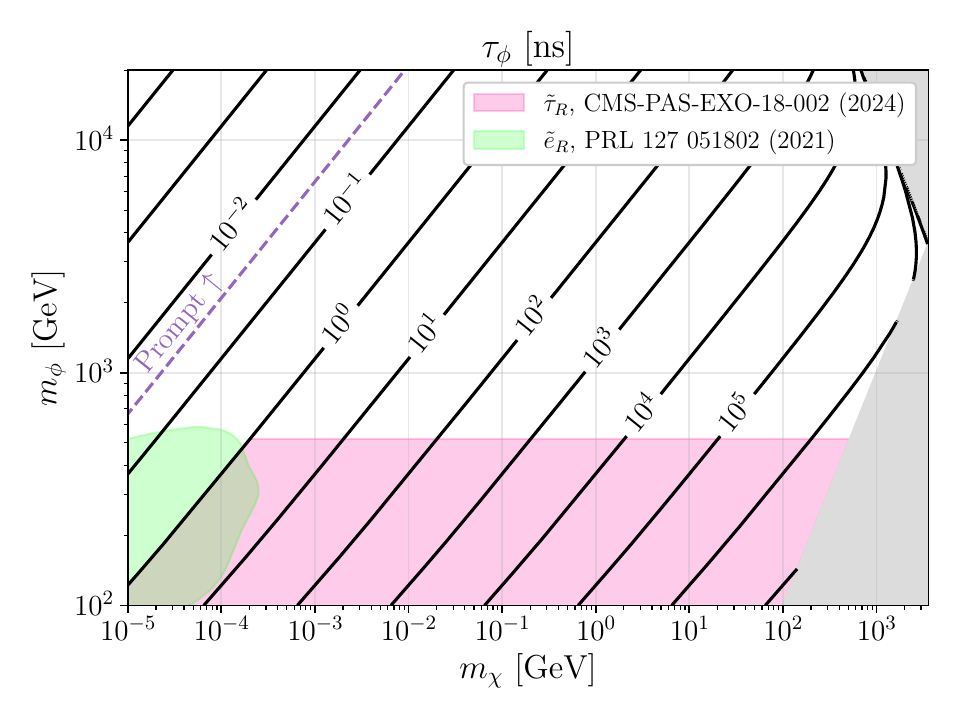}
    }
    \caption{The charged mediator $\phi$ proper lifetime, $\tau_\phi$, assuming the Yukawa couplings required to get the right relic abundance (as given in Figure~\ref{fig:lambda_parameter}). Depending on this quantity, the signal, and thus the collider search strategy, varies significantly across the parameter space. The green (pink) region is already ruled out by the LHC search in Ref.~\cite{ATLAS:2020wjh} (Ref.~\cite{CMS:2024qys}). This figure shows a large part of the viable parameter space is not probed by the LHC, motivating future collider searches. The dashed purple line denotes the boundary between our prompt (Section~\ref{sec:prompt}) and LLP (Section~\ref{sec:displaced}) search regions, see Section~\ref{sec:main} for details. }
    \label{fig:lifetime}
\end{figure}

The on-going searches at LHC already constrain parts of our parameter space. These constraints depend strongly on the lifetime of the mediator. 
In Figure~\ref{fig:lifetime} we show the lifetime of the mediator as given by
\begin{equation}
\tau_\phi = (3 \Gamma_{\phi \to \ell \chi})^{-1},
    \label{eq:lifetime}
\end{equation}
where the factor of three is due to the three charged lepton channels. 
In this calculation we assume the same branching ratio to each final state lepton. 
This lifetime affects the signal of our model at future colliders as well as at the LHC. 

To find the bounds from LHC, we note that our mediator is identical to a right-handed slepton. 
Many searches for sleptons have been carried out at the LHC, \textit{e.g.} see \cite{ATLAS:2022hbt, ATLAS:2019lff}, however, the majority of these searches assume that the sleptons decay promptly. To map the prompt region onto our parameter space, we take anything that has a proper lifetime of $\tau_\phi \lesssim 3 \times 10^{-2}$~ns as prompt (see further details in section \ref{subsec:basics_dm_at_muc}). With this, the prompt searches do not reach high enough $\phi$ masses to be visible on our bounds plot, regardless of combination of slepton flavors. 

On the other hand, searches for long-lived sleptons can probe some parts of the parameter space.
In Figure~\ref{fig:lifetime} we show the bounds from the long-lived sleptons in Ref.~\cite{ATLAS:2020wjh} on our model. 
In applying these bounds we assumed a branching ratio of 1 to the strongest charged lepton channel (selectrons in this case), to be optimistic about the reach of LHC in our parameter space. 
We also show the bounds from CMS search for staus as heavy stable charged particles in the detector \cite{CMS:2024qys}. 
Figure \ref{fig:lifetime} clearly shows a substantial part of the viable parameter space remains unavailable to LHC, and this motivates searches for this model in future colliders, which we turn to next.
Additionally, limits from LEP are always taken into account by assuming that $m_\phi > 100$ GeV \cite{DELPHI:2003uqw,ALEPH:2002gap} and small-scale cosmological structure observables set a limit on the mass of freeze-in DM such that $m_\chi \gtrsim 10$ keV \cite{Garny:2018ali,DEramo:2020gpr,Decant:2021mhj}.

\section{Production at Muon Colliders}
\label{sec:main}

We now turn to signals of our flavored DM model at a MuC.  
In our analysis we assume there exists only one flavor of DM. 
We also remain inclusive over the final state charged leptons in our analysis. 
Our mediator production rate is independent of the branching ratio to different charged leptons and can be repeated straightforwardly for models where the mediator decays with different branching ratios. 

We will calculate the signal yield of our model at a future MuC and comment on the reach of a future MuC in its parameter space.
When possible, we will also report the tolerable systematic uncertainty that will still allow a discovery of our model with the aim that our results can inform the detector designs and simulations, especially with regards to LLP signals. 
We start with a review of the relevant aspects of a future MuC design and then discuss the production cross-section of our model's mediator at such a facility.

\subsection{A Future Muon Collider}
\label{subsec:muc}

There are many on-going experimental studies about the feasibility and features of a future MuC, including works on a detector design; see Refs.~\cite{MuonCollider:2022xlm,MuonCollider:2022glg,MuonCollider:2022nsa,MuonCollider:2022ded,Black:2022cth,Accettura:2023ked} for recent reviews. 
A final and complete detector design is still an on-going field of research. 
While this prevents a detailed study of phenomenology of BSM models at a MuC, it underscores the importance of developing a template of canonical models that, in turn, informs the development of the detector.

While further studies are in order, recent progress in the cooling system \cite{MICE:2019jkl} has revived the hopes for achieving large luminosities at a MuC.
The benchmark target luminosity for a high energy MuC with center of mass energy of $\sqrt{s}$ is given by \cite{Delahaye:2019omf}
\begin{equation}
    \mathcal{L} \approx 10 ~ \mathrm{ab}^{-1} \times \left(      \frac{\sqrt{s}}{10~\mathrm{TeV}} \right)^2,
    \label{eq:Lumin}
\end{equation}
which we use in our study as well. We will focus on bechmark center of mass energies $\sqrt{s} = 3$~TeV and $\sqrt{s} = 10$~TeV, for which Eq.~\eqref{eq:Lumin} suggests benchmark $1$ ($10$) ab$^{-1}$ integrated luminosities, respectively.

In Section~\ref{sec:displaced} we outline an LLP search for our model, whose primary signature is displaced leptons at different transverse distances from the beam. 
This highlights the need for a rough estimate for the size of different components of the detector, hence the need for assuming a benchmark detector design.
We use the design sketched in Ref.~\cite{MuonCollider:2022ded,Black:2022cth}, which in turn is informed by the existing lepton collider designs, \textit{e.g.} see the CLIC \cite{CLICdp:2017vju} and ILC designs \cite{ILC:2007vrf}. The size of different segments of the detector are reported in Table~\ref{tab:detector} (table taken from Ref.~\cite{MuonCollider:2022ded}).
We will use the barrel region of these segments in our LLP analysis of Section~\ref{sec:displaced}. 

\begin{table}
    \centering
    \resizebox{\columnwidth}{!}{
    \begin{tabular}{r|l||c|c|c|l}
        \textbf{Subsystem} & \textbf{Region} & \textbf{$L$ dimensions [cm]} & \textbf{$|Z|$ dimensions [cm]} & $\mathbf{\eta}$ \textbf{bound} & \textbf{Material} \\
        \hline\hline
        Vertex Detector & Barrel & $3.0 - 10.4$     & $65.0$    & $\lesssim 2.53$ 
        & Si \\
                        & Endcap & $2.5 - 11.2$     & $8.0 - 28.2$  & $\lesssim 1.65$ 
                        & Si \\
        \hline
        Inner Tracker   & Barrel & $12.7 - 55.4$    & $48.2 - 69.2$  & $\lesssim 1.05$ 
        & Si \\
                        & Endcap & $40.5 - 55.5$    & $52.4 - 219.0$  & $\lesssim 1.07$ 
                        & Si \\
        \hline
        Outer Tracker   & Barrel & $81.9 - 148.6$   & $124.9$     & $\lesssim 0.76$ 
        & Si \\
                        & Endcap & $61.8 - 143.0$   & $131.0 - 219.0$ & $\lesssim 1.21$ 
                        & Si \\
        \hline\hline
        ECAL            & Barrel & $150.0 - 170.2$  & $221.0$      & $\lesssim 1.08$ 
        & W + Si \\
                        & Endcap & $31.0 - 170.0$   & $230.7 - 250.9$ & $\lesssim 1.18$ 
                        & W + Si \\
        \hline
        HCAL            & Barrel & $174.0 - 333.0$  & $221.0$      & $\lesssim 0.62$ 
        & Fe + PS \\
                        & Endcap & $307.0 - 324.6$  & $235.4 - 412.9$ & $\lesssim 0.71$ 
                        & Fe + PS \\
        \hline\hline
        Solenoid        & Barrel & $348.3 - 429.0$  & $412.9$        & $\lesssim 0.85$ 
        & Al \\
        \hline\hline
        Muon Detector   & Barrel & $446.1 - 645.0$  & $417.9$     &   $\lesssim 0.61$ 
        & Fe + RPC \\
                        & Endcap & $57.5 - 645.0$   & $417.9 - 563.8$  & $\lesssim 0.79$ 
                        & Fe + RPC \\
    \end{tabular}
    }
    \caption{Different components of the current proposed detector design for a high energy MuC.  Table is adapted from Ref.~\cite{MuonCollider:2022ded}. The $L$ ($Z$) dimension refers to the segment size in the transverse (longitudinal) direction. The $\eta$ bound column is added to the table of Ref.~\cite{MuonCollider:2022ded} and shows the highest value of pseudorapidity $\eta$ for a track that completely goes through that region. This information will enter our LLP search in Section~\ref{sec:displaced}. }
    \label{tab:detector}
\end{table}

The closest part of the detector to the beam is a vertex detector that spans transverse distances of up to ten centimeters. 
The next part is the silicon-based tracking system, filling up the space until transverse distance of around a meter.
Outside the tracking detector reside the silicon-tungsten ECAL with thickness of tens of centimeters and a multi-layer HCAL, followed by the magnetic field solenoids stretching roughly up to a radius of four meters away from the beam.
Finally, a muon spectroscopy system starts at around transverse distances of four meters and surrounds the entire system.

In addition to these, to mitigate the effect of the beam-induced background 
(BIB) at a MuC, two tungsten nozzles cladded with borated polyethylene are included in each forward direction. These nozzles limit the access to forward directions; we will assume pseudorapidity $|\eta| \gtrsim 2.4$ is not accessible. See Refs.~\cite{ILC:2007vrf,MuonCollider:2022ded,Black:2022cth} for further details about the design. 

We should remind the reader that Table~\ref{tab:detector} should not be construed as the final MuC detector design. 
Our upcoming analysis can straightforwardly be repeated once a concrete design is developed.

The design in Ref.~\cite{MuonCollider:2022ded} uses a large magnetic field as well. Nonetheless, the effect of this magnetic field will be negligible on our study. 
In particular, our proposed LLP search in Section~\ref{sec:displaced} relies on transverse displacement of displaced leptons, which is not particularly sensitive to the magnetic field (for any feasible magnetic field size in the detector).\footnote{An adequately precise measurement of some LLP signals, such as closest approach of the track to the primary vertex, requires a proper inclusion of the magnetic field effect - see Ref.~\cite{Knapen:2022afb} for a recent review. However, we will not study such signals in this paper. 
}

\subsection{Flavored Dark Matter at a Future Muon Collider}
\label{subsec:basics_dm_at_muc}

In this section we go over basics of our model signals in a future MuC. 
Since we are focusing on the freeze-in regime, the Yukawa coupling of the DM to SM leptons and the mediator $\phi$ is very small, thus direct DM production is strongly suppressed. 
The mediator $\phi$, on the other hand, is charged under SM gauge groups and can be produced ubiquitously if kinematically allowed. 
Subsequently, $\phi$ can only decay to the DM and a lepton, giving rise to a charged lepton and missing energy in the detector. 
Since the production is determined by SM gauge couplings, our model is a well-motivated target whose signal yield can be used for a plethora of different BSM models, and can inform future detector designs. 

\begin{figure}
    \centering
    \resizebox{0.35\columnwidth}{!}{
    \includegraphics{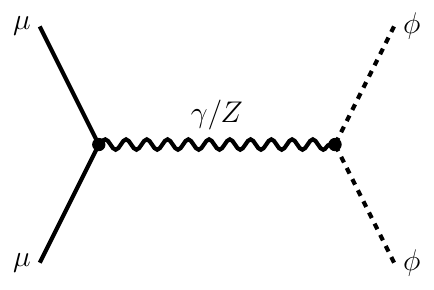}
    }\\
    \resizebox{\columnwidth}{!}{
    \includegraphics{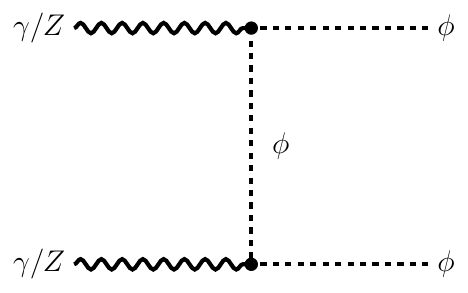}
    \hspace{0.2cm}
    \includegraphics{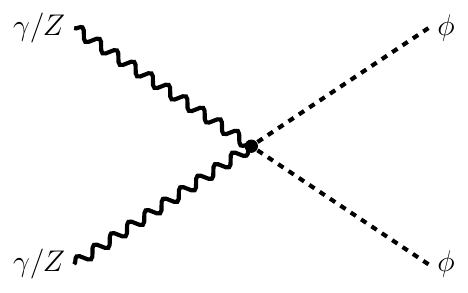}
    \hspace{0.2cm}
    \includegraphics{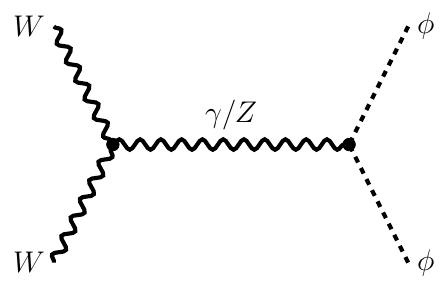}
    }
    \caption{Diagrams pair-producing $\phi$ at a future MuC. The initial state can be either SM charged leptons (\textbf{top}), or electroweak gauge bosons (\textbf{bottom}). We refer to these production channels as Drell-Yan (DY) or vector boson fusion (VBF), respectively. The produced $\phi$ particles are on-shell and will eventually decay to DM $\chi$ and a charged lepton. All diagrams involving an off-shell $\phi$ or their single-production are suppressed by the small Yukawa coupling to DM in our freeze-in setup.}
    \label{fig:collider_diagrams}
\end{figure}

Diagrams relevant for $\phi$ production at a MuC are shown in Figure~\ref{fig:collider_diagrams}. The kinematic distribution of final states will change depending on the production channel.
Similar topologies exist for other flavored DM setups (where $\phi$ has same charges as other SM fermions). They give rise to a pair of $\phi$ particles, each of which subsequently decays to a DM particle and a lepton. 
The branching ratio to different charge leptons will depend on the Yukawa couplings. Without any further structure on the flavor texture, we expect these different branching ratios to be comparable. We remain agnostic about the value of these branching ratios in this work.

The top diagram in Figure~\ref{fig:collider_diagrams}, which resembles a Drell-Yan (DY) production, comes from initial state muons, whose PDF \cite{Han:2020uid,Han:2021kes,Ruiz:2021tdt} is dominated by longitudinal fractional momentum $x = 1$, while the other three diagrams are similar to vector boson fusion (VBF) production channels. 
In the VBF production channels the original lepton typically goes undetected in the forward direction down the beam. Nonetheless, it sometimes recoils enough to fall within the detectable $\eta$ range and will give rise to an extra charged lepton in the final state. We will not consider such events in our search.

\begin{figure}
    \centering
    \resizebox{0.8\columnwidth}{!}{
    \includegraphics{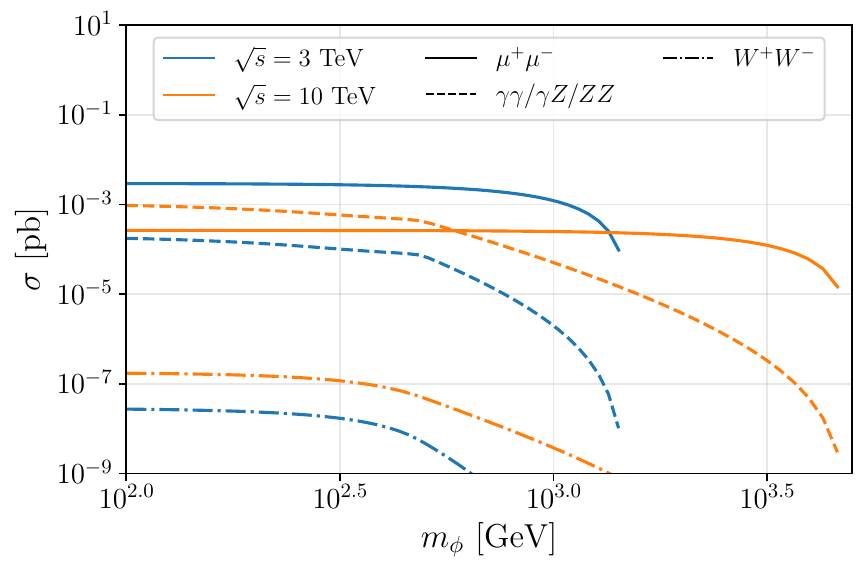}
    }
    \caption{Cross-section for pair production of $\phi$ mediators at a future MuC with center of mass energy of 3~TeV (\textbf{blue}) or 10~TeV (\textbf{orange}). We do this calculation using \texttt{MadGraph5}~\cite{Alwall:2011uj}. Results for different initial states are shown with line styles, with DY as solid, $W$ VBF as dot-dashed, and $\gamma/Z$ VBF as dashed. We find that the production is dominated by the DY channel for the mediator masses above a few hundred GeV.}
    \label{fig:cross_sections}
\end{figure}

We use \texttt{MadGraph5} \cite{Alwall:2011uj} for event generation and study of kinematic distributions. 
Based on the existing results on various SM particles PDF in a muon \cite{Han:2020uid,Han:2021kes,Ruiz:2021tdt}, in our simulations we assume the muons are all produced with longitudinal fractional momentum $x=1$, while electroweak gauge bosons are modeled with the effective vector approximation (EVA) and already included in \texttt{MadGraph5} \cite{Ruiz:2021tdt}. 
The total cross-section for $\phi$ pair production, from various initial states are shown in Figure~\ref{fig:cross_sections}. 
We can see from these figures that for heavy mediator masses the production is strongly dominated by DY production channels, while at low $m_\phi$ values the VBF channel contribution become more relevant (especially for center of mass energy $\sqrt{s}=10$~TeV).
The contribution from DY processes with other SM fermion initial states are negligible compared to the aforementioned production channels.

The model's signal at a collider is substantially affected by the lifetime of the $\phi$ mediator, see Figure~\ref{fig:lifetime}. 
We divide the parameter space of the model by $\phi$ lifetime, $\tau_\phi$. Specifically, we will consider two different strategies. 
\begin{itemize}
    \item For $\tau_\phi \lesssim 3 \times 10^{-2}$~ns the mediator decays \textbf{promptly} to a lepton and the missing energy in the vicinity of the interaction point.

    \item For $ \tau_\phi \gtrsim 3 \times 10^{-2}~\mathrm{ns}$ the produced mediator particles are \textbf{LLP}s, moving inside the detector for a macroscopic distance and leaving a charged track. Depending on the lifetime, they can either decay inside the detector (giving rise to displaced leptons), or escape it (like heavy stable charged particles). In either case, they will leave a detectable charged track in some parts of the detector. 
\end{itemize}

It should be emphasized that our division of the parameter space into prompt and LLP is dependent on detector geometry and exact process details; whether a cut-and-count analysis for prompt signals is applicable to longer lifetimes, or if an LLP search strategy to shorter lifetimes, should be studied on a case-by-case basis and once a final detector design exists.

In the upcoming sections we detail a search strategy and calculate the signal yield in each region. 
We should emphasize that our proposal is a rudimentary analysis intended to lay out simple search strategies and to demonstrate the power of a high enregy MuC in probing this well-motivated model. 
In the prompt region we calculate the SM background and report the systematics uncertainties that can be tolerated while still allowing for a discovery. 
In the LLP region the irreducible background is generated via BIB and the detector response, better understanding of which requires simulations beyond the current work. Thus, in this region we simply report the signal yield and some of its unique features, leaving further studies of the background, systematics, and the reach of a high energy MuC to future works.

\section{A Search for the Prompt Region}
\label{sec:prompt}

\subsection{Kinematics and Signal Regions}
\label{subsec:prompt_srs}

As the name implies, in this part of the parameter space the signal of our model will be a pair of charged leptons and missing energy coming from the prompt decay of the mediator $\phi$.  
In this section we go over the kinematics of events in this part of the parameter space and put forward a rudimentary search proposal. We will show that our proposal can discover our model almost over the entire kinematically-accessible region.

\begin{figure}
    \centering
    \resizebox{0.8\columnwidth}{!}{
    \includegraphics{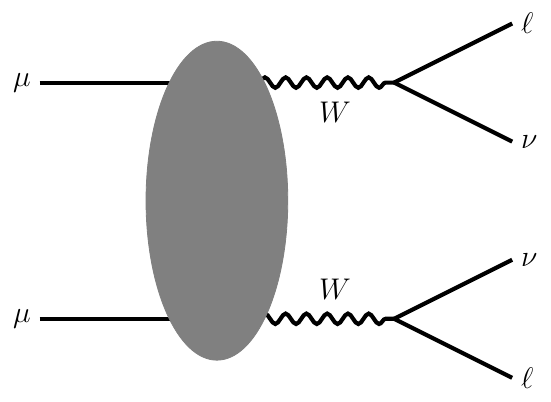}
    \hspace{3cm}
    \includegraphics{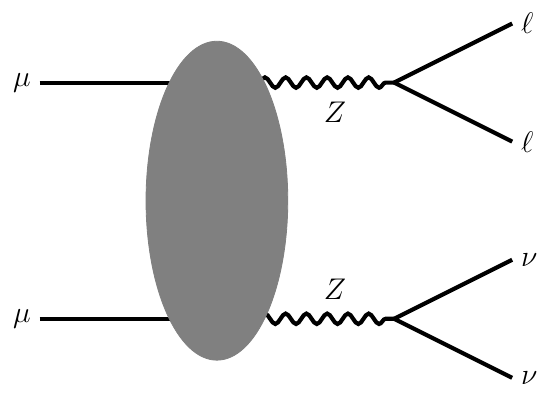}
    }
    \caption{Dominant SM background for our search in the prompt region. The $W^+W^-$ pair-production is potentially the most constraining background; we can suppress this background by applying $M_{T2}$ cuts, see the text for further details.}
    \label{fig:bkg_prompt}
\end{figure}

The cross-section for $\phi$ pair production, from various initial states are shown in Figure~\ref{fig:cross_sections}. 
The main background for this search comes from electroweak gauge bosons pair production and decay in Figure~\ref{fig:bkg_prompt}.

To reduce these backgrounds, we propose a cut on the event's transverse missing mass $M_{T2}$ \cite{Lester:1999tx}, the lepton system invariant mass $m_{\ell\ell}$, and the lepton system transverse momentum $p_{T,\ell\ell}$. 
The background from a pair production of $Z$ bosons can be significantly reduced by $m_{\ell\ell}$ or $p_{T,\ell\ell}$.
This is due to the fact that charged leptons originating from SM gauge bosons decays have a much lower invariant mass $m_{\ell\ell}$ and transverse momentum $p_{T,\ell\ell}$ compared to leptons coming from our mediator's decay. 
The remaining background from pair production of the $W$ boson can also be strongly suppressed by the use of $M_{T2}$. 
To see that, we should keep in mind that in our pair production topologies $M_{T2}$ is bounded from above by the mediator mass \cite{Lester:1999tx}. Thus, $W$ pair production events have much lower $M_{T2}$ values than the signal events from decays of $\phi$.

\begin{figure}
    \centering
    \resizebox{\columnwidth}{!}{
    \includegraphics{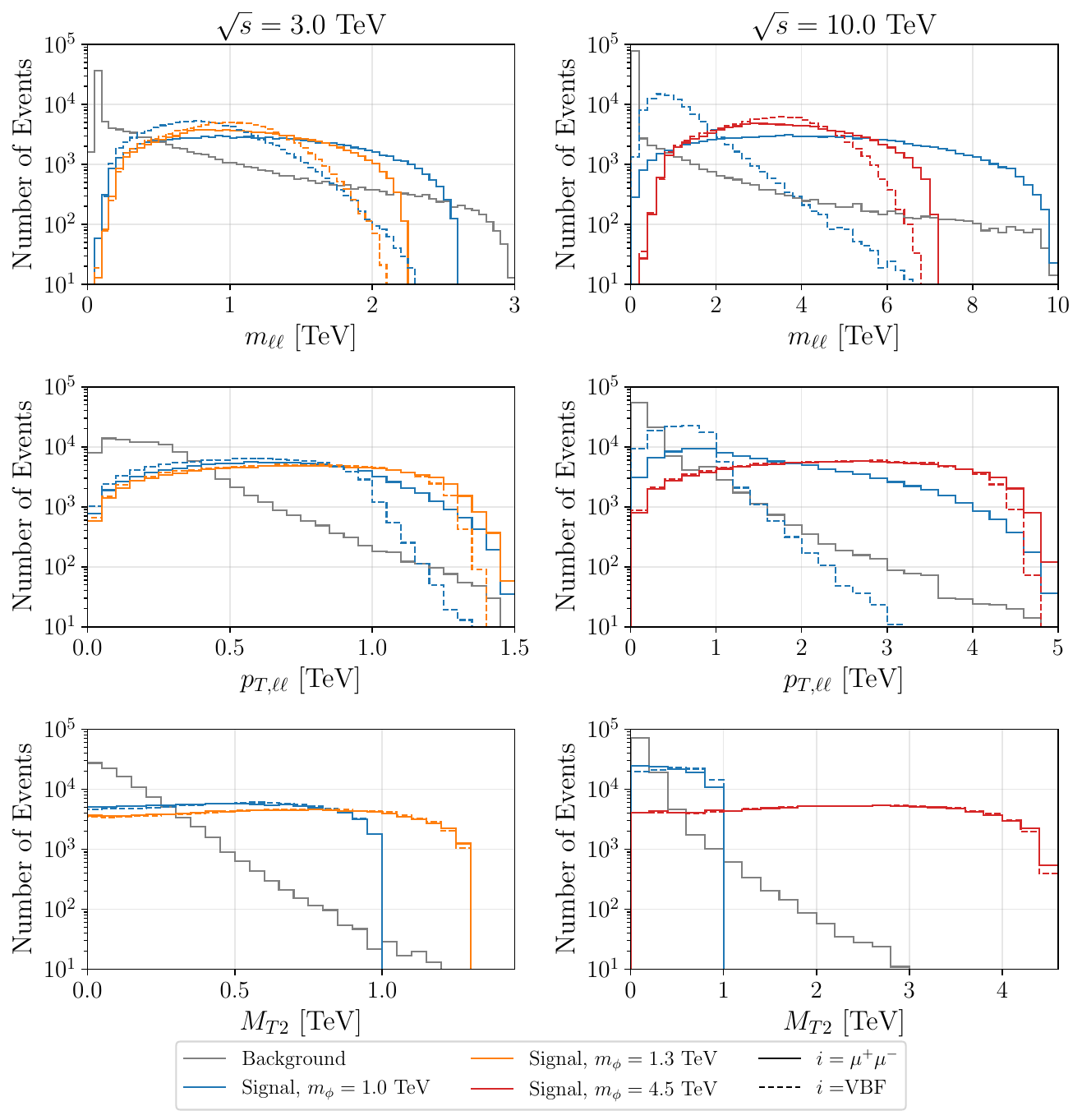}
    }
    \caption{Histograms of invariant lepton mass (\textbf{top}), lepton transverse momentum (\textbf{middle}), and $M_{T2}$ (\textbf{bottom}) for our model (colored) and for SM (gray) at a MuC with center of mass energies of 3~TeV (\textbf{left}) and 10~TeV (\textbf{right}). The bin size for each observable is 50 (200) GeV for $\sqrt{s} = 3$ $(10)$ TeV.
    We use \texttt{MadGraph5} with $10^5$ events to generate each histogram.
    The solid (dashed) colored histograms refer to signal processes from the DY (VBF) channel. 
    In each figure we show the distribution for two $\phi$ masses towards the two kinematically accessible ends of the mass ranges. We find a sharp contrast between the SM's and our model's predictions, motivating use of these cuts in our analysis. }
    \label{fig:histograms_1D_prompt}
\end{figure}

To better justify using these cuts, in Figure~\ref{fig:histograms_1D_prompt} we show the distribution of events in various kinematic observables, as well as the SM background distributions. 
We find that the distributions of the SM background and our model differ by orders of magnitude in certain ranges of $M_{T2}$, $m_{\ell\ell}$, and $p_{T,\ell\ell}$, hence motivating use of these variables.
In particular, we find SM background distributions are concentrated at lower values of these parameters, thus justifying the use of a lower bound cut on their value in our analysis. 
We should also note the difference in the distribution of events from the two signal channels (DY and VBF - see Figure~\ref{fig:collider_diagrams}).

We use \texttt{MadGraph5}~\cite{Alwall:2011uj} to generate events for these histograms. 
For SM background, we consider the processes $\mu^- \mu^+ \rightarrow \ell^- \ell'^+ \bar{\nu}_\ell \nu_{\ell'}$ with $\ell^{(')} \in [e, \mu]$. 
Other background channels that come from VBF diagrams (namely $ V V \rightarrow \mu \mu \nu \nu$) can be approximated with the Improved Weizsacker-Williams (IWW) \cite{PhysRevD.8.3109,Frixione:1993yw} implementation in \texttt{MadGraph5}. We find that the VBF contributions give rise to sub-percent level corrections to the background and can be safely ignored, even before any kinematic cuts.

We propose putting a cut on the three kinematic variables $M_{T2}$, $m_{\ell\ell}$, and $p_{T,\ell\ell}$ and counting the number of events from our model and from SM backgrounds. We propose different signal regions (with different cuts on each kinematics observable) and allow for different signal regions to have overlaps. For every point in the parameter space we choose the signal region that has the most sensitivity to our signal.

Assuming a Gaussian distribution of SM background events in each signal region, the statistical error bar on each signal region is simply square root of the average number of SM events in that region, denoted by $\sqrt{B}$. 
Hence, we can quantify the significance of the signal count $S$ in each signal region by simply calculating $S/\sqrt{B}$.
(For signal regions for which our \texttt{MadGraph5} simulations predict $\sqrt{B} \leqslant 2$, we use statistical error of $\sqrt{B}=2$ instead to be conservative.) 
For any given mediator mass, all regions with $S/\sqrt{B} \geqslant 5$ allow a $5\sigma$ discovery of our model.\footnote{Since we use individual signal regions for a discovery, we do not need to resort to more sophisticated likelihood calculations; repeating this analysis with more complete statistical treatment is straightforward. }

For $\sqrt{s}=10$~TeV, we find the cuts on $m_{\ell\ell}$, $p_{T,\ell\ell}$, and $M_{T2}$ that optimize our sensitivity to the model with $m_\phi = 1,4.5$~TeV: 
\begin{eqnarray}
    \label{eq:cuts_optimize1_10}
    \mathrm{Optimized~Sensitivity~} m_\phi = 1~\mathrm{TeV}~:~(m_{\ell\ell},p_{T,\ell\ell},M_{T2}) = (3.60,1.30,0.20)~\mathrm{TeV} \\
    \label{eq:cuts_optimize4_10}
    \mathrm{Optimized~Sensitivity~} m_\phi = 4.5~\mathrm{TeV}~:~(m_{\ell\ell},p_{T,\ell\ell},M_{T2}) = (1.50,0.20,3.50)~\mathrm{TeV}
\end{eqnarray}
For $\sqrt{s}=3$~TeV, we find the cuts on $m_{\ell\ell}$, $p_{T,\ell\ell}$, and $M_{T2}$ that optimize our sensitivity to the model with $m_\phi = 1,1.3$~TeV
\begin{eqnarray}
    \label{eq:cuts_optimize1_3}
    \mathrm{Optimized~Sensitivity~} m_\phi = 1~\mathrm{TeV}~:~(m_{\ell\ell},p_{T,\ell\ell},M_{T2}) = (0.45,0.69,0.53)~\mathrm{TeV} \\
    \label{eq:cuts_optimize1.4_3}
    \mathrm{Optimized~Sensitivity~} m_\phi = 1.3~\mathrm{TeV}~:~(m_{\ell\ell},p_{T,\ell\ell},M_{T2}) = (0.26,0.87,0.86)~\mathrm{TeV}
\end{eqnarray}
Further details about how we arrive at these cuts are included in Appendix~\ref{app:prompt}. 
We use the benchmark mass $m_\phi = 1$~TeV since it is around the lowest mediator mass inside the prompt region, see Figure~\ref{fig:lifetime}; $m_\phi = 4.5$~TeV ($m_\phi = 1.3$~TeV) for $\sqrt{s}=10$~TeV ($\sqrt{s}=3$~TeV) is also around the maximum mass ($\sqrt{s}/2$) that is kinematically accessible. 
We should note that since the mediator coupling to DM is very small, the signal is completely suppressed for the mass ranges where the mediator can not be produced on-shell.

We use the cuts from Eqs.~\eqref{eq:cuts_optimize1_10}--\eqref{eq:cuts_optimize1.4_3} and the average of the cuts on the two extreme mass cases for each center of mass energy to define 27 overlapping signal regions for each center of mass energies. 
For each point in the parameter space we identify the signal region that has the best $S/\sqrt{B}$.
In Figure~\ref{fig:promptSRs} we show signal regions, as defined in Appendix~\ref{app:prompt}, that are most sensitive to each point in the parameter space and allow a $5\sigma$ discovery. 
The marked mass range for each signal region shows where that signal region gives rise to a discovery, while neglecting systematics; the small dots for each mass point indicate the best-fit signal region for that mediator mass. 
It should be noted that this analysis is independent of the DM mass $m_\chi$; for virtually every point in the prompt decay region the DM can be treated as massless.

\begin{figure}
    \centering
    \resizebox{0.9\columnwidth}{!}{
    \includegraphics{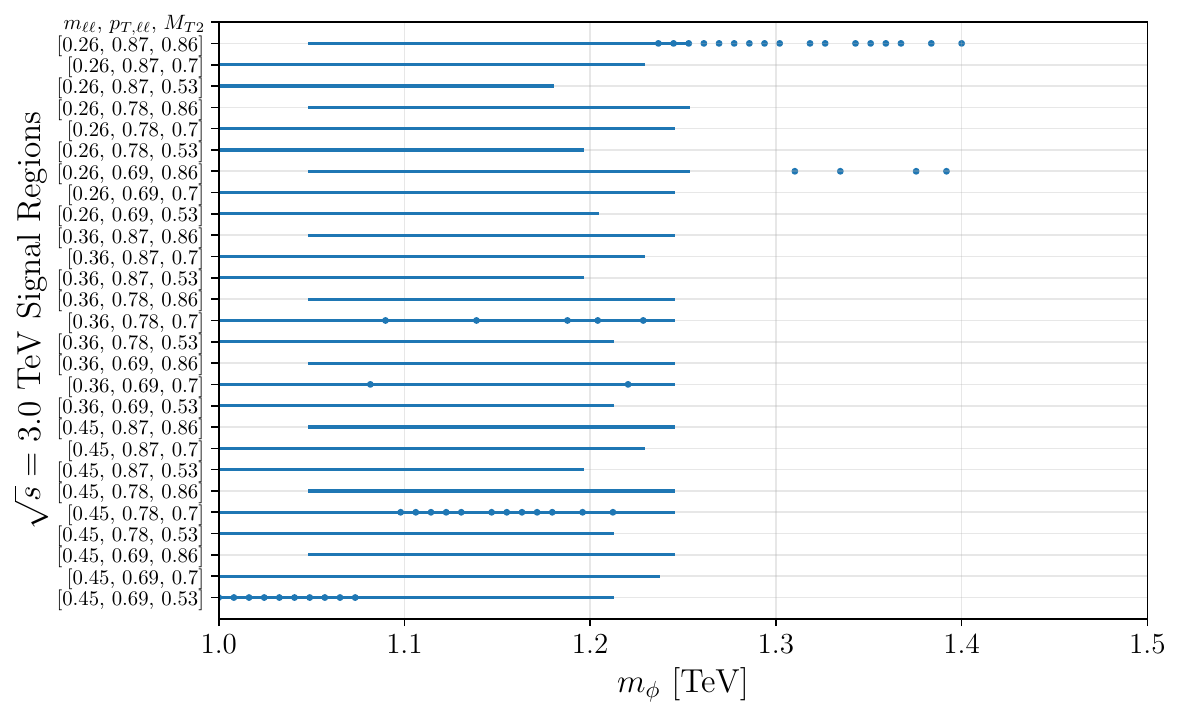}
    }\\
    \resizebox{0.9\columnwidth}{!}{
    \includegraphics{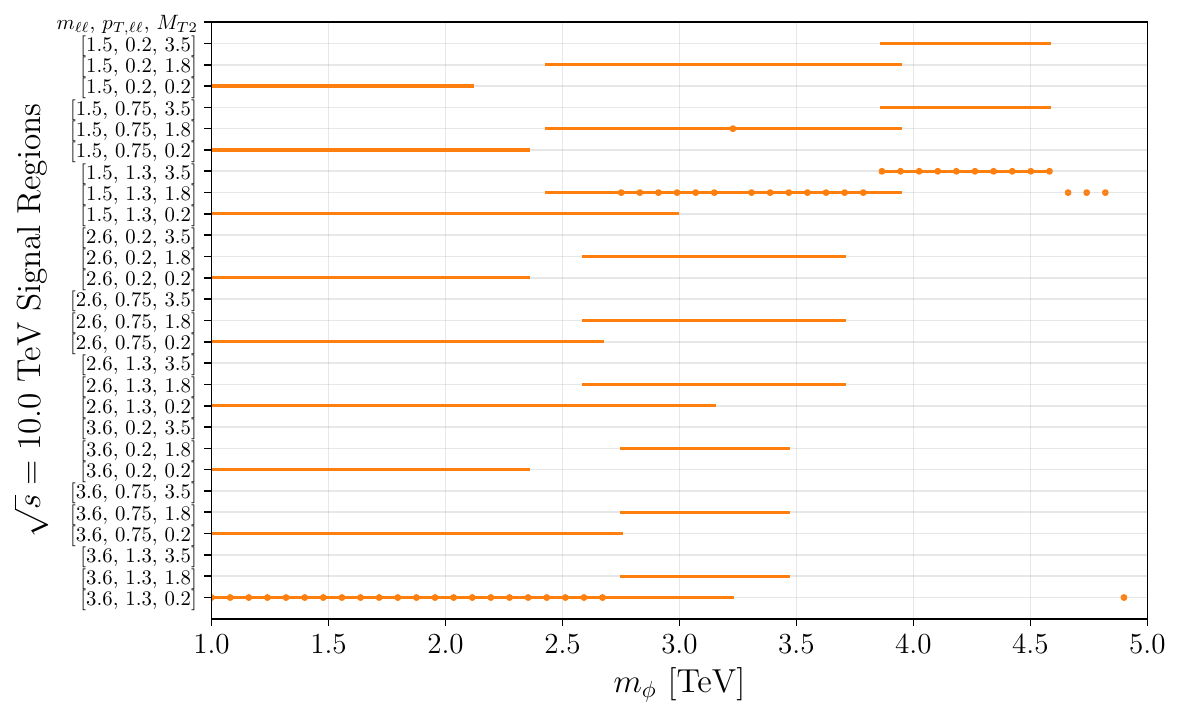}
    }
    \caption{The range of mediator masses that each signal region could discover with $5\sigma$ confidence at a future MuC with center of mass energies of 3~TeV (\textbf{top}) or 10~TeV (\textbf{bottom}), assuming zero systematic uncertainties. The lines indicate the mass range in which each signal region yields a discovery, while the dots denote the best-fit signal region for each mass point. We find that for all masses above a TeV, and almost upto $\sqrt{s}/2$, a future MuC can discover the signal of our model. }
    \label{fig:promptSRs}
\end{figure}

We find that the cuts are strong enough that essentially any part of the parameter space where $\phi$ pairs can be produced on-shell can be probed in future MuCs. 
We find that for most mediator masses there are multiple signal regions that allow a $5\sigma$ discovery.

\subsection{Target Systematics}
\label{subsec:prompt_sys}

We have thus far ignored various systematics uncertainties for each signal region. 
There are many on-going experimental efforts for designing the MuC detector and estimation of systematic uncertainties. Tolerable systematics on well-motivated models, such as the ones studied in this work, can serve as a motivated target in such studies. 

\begin{figure}
    \centering
    \resizebox{1.\columnwidth}{!}{
    \includegraphics{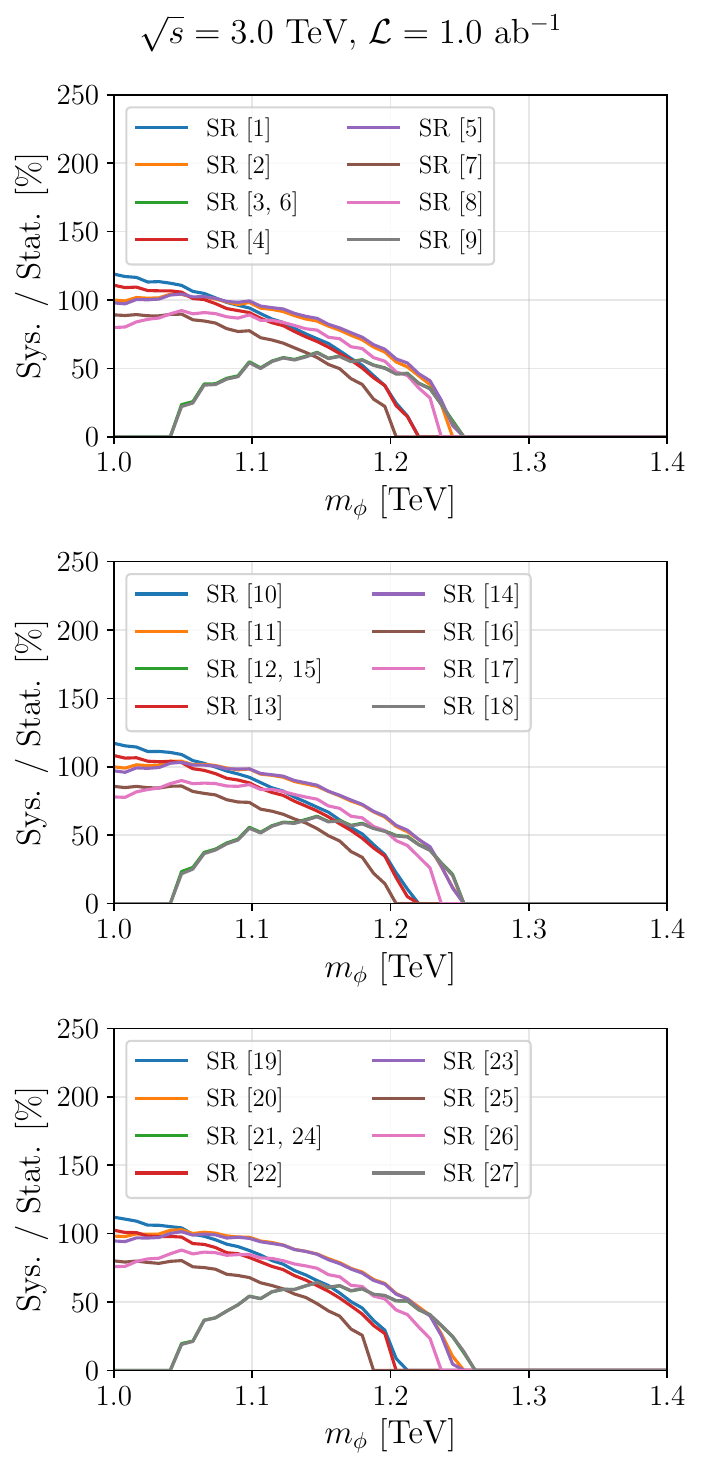}
    \includegraphics{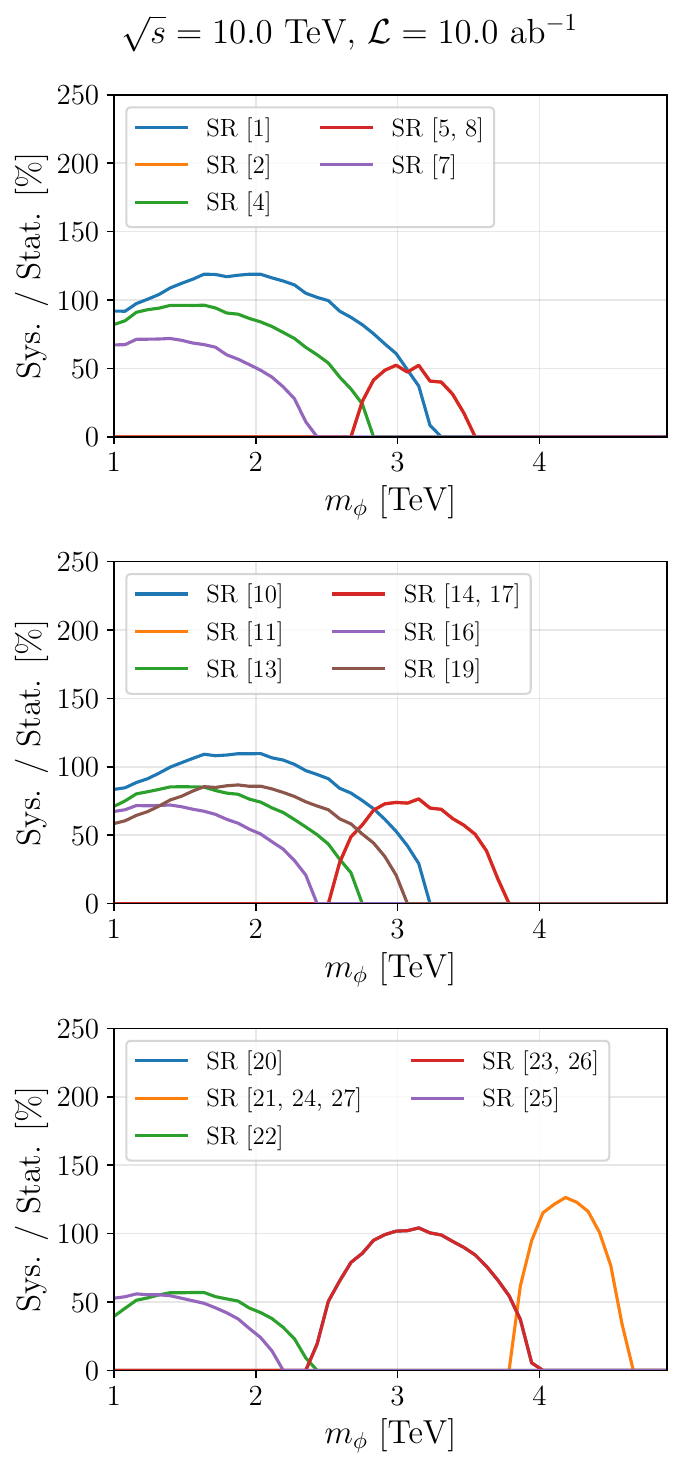}
    }
    \caption{Tolerable systematic uncertainty for each signal region, as a fraction of statistical uncertainty, that still allows a discovery of our model for different mediator masses, see Figure.~\ref{fig:SRdefs} for definitions of signal regions. The \textbf{left} (\textbf{right}) column includes signal regions at a 3 TeV (10 TeV) MuC. We only include signal regions with more than twenty signal events. Signal regions not included in this figure either have fewer signal events or do not give rise to a $5\sigma$ discovery of the signal, even in the absence of systematics uncertainties. }
    \label{fig:money_plot}
\end{figure}

In Figure~\ref{fig:money_plot} for each $\phi$ mass we calculate the maximum systematic uncertainties (as a fraction of the statistical uncertainty in that region) that still allows a discovery of our model for that point in the parameter space. 
In doing so, we assume systematics and statistical uncertainties are uncorrelated. 
(We also used $\sqrt{B}=2$ for signal regions that have $\sqrt{B} \leqslant 2$ in our simulations.)

This figure shows that even with systematics uncertainties that are comparable to the statistical error in the most-sensitive signal regions, we can still discover our model in the kinematically accessible parts of its parameter space.

Finally, we should point out that in our analysis we used the average background prediction in each signal region according to the results of \texttt{MadGraph5} simulations. 
Theoretical uncertainties on this prediction should also be included in an actual analysis by smearing the SM prediction in each signal region by a Gaussian distribution. 
We will leave a thorough treatment of all systematics, statistical, and theoretical uncertainties for future works.

It should also be noted that our search can be repeated for other types of flavored DM models. In the case of quark flavored models, we expect to have a jet, instead of the charged lepton, from the primary vertex and our search should be updated accordingly. We expect cuts on similar set of kinematics observables will allow us to probe the prompt region parameter space of all such models as well.

\section{A Search for the Long-Lived Region}
\label{sec:displaced}

For longer lifetimes, the pair of mediators can leave a charged track in the detector and give rise to various LLP signals, see Refs.~\cite{Lee:2018pag,Knapen:2022afb} for recent reviews of LLPs. 
In our setup, depending on the lifetime, the produced $\phi$ particles subsequently either decay to a charged lepton and a DM particle inside the detector, or escape the detector without decaying. 
Further studies are required for better understanding of the background for these searches, see the discussion below. 
This prevents us from reporting the discovery reach of a future MuC in our model's parameter space at the moment.
Instead, here we focus on the signal yield of our model. In particular, we will calculate the distribution of expected number of displaced leptons in different detector components.\footnote{In our model the mediators are always produced in pairs. Thus, each signal event includes two charged tracks and two displaced leptons that could further enhance the search \cite{Buchmueller:2017uqu}. We, however, will not explicitly use this property in our rudimentary analysis here.} 
We also show that our model exhibits a double-peak feature in some kinematic distributions, that distinguishes it from SM background. 
Given the fact that our setup is a theoretically-motivated target and its production rate and signals are generic for any electroweakly-charged LLP, we hope this signal yield can serve as a target for experimental studies and detector design efforts.

\subsection{Long-Lived Particle Kinematics}
\label{subsec:LLP_kinematics}

Previous studies of LLPs at a high energy MuC focused on scenarios where the long lifetime is an artifact of a small mass splitting between the LLP and its neutral daughter \cite{Capdevilla:2021fmj,Bottaro:2021snn,Bottaro:2022one}. 
In such scenarios the charged decay product, \textit{e.g.} a SM charged lepton, can carry a small momentum and escapes detection, giving rise to a disappearing track signature. 
In our setup, on the other hand, the mediator and DM have a large mass splitting and the long lifetime of the mediator is an artifact of a small Yukawa coupling. 
As a result of this large mass splitting, if the mediator decays inside the detector the charged lepton will be easily detectable.

\begin{figure}
    \centering
    \resizebox{1.\columnwidth}{!}{
    \includegraphics{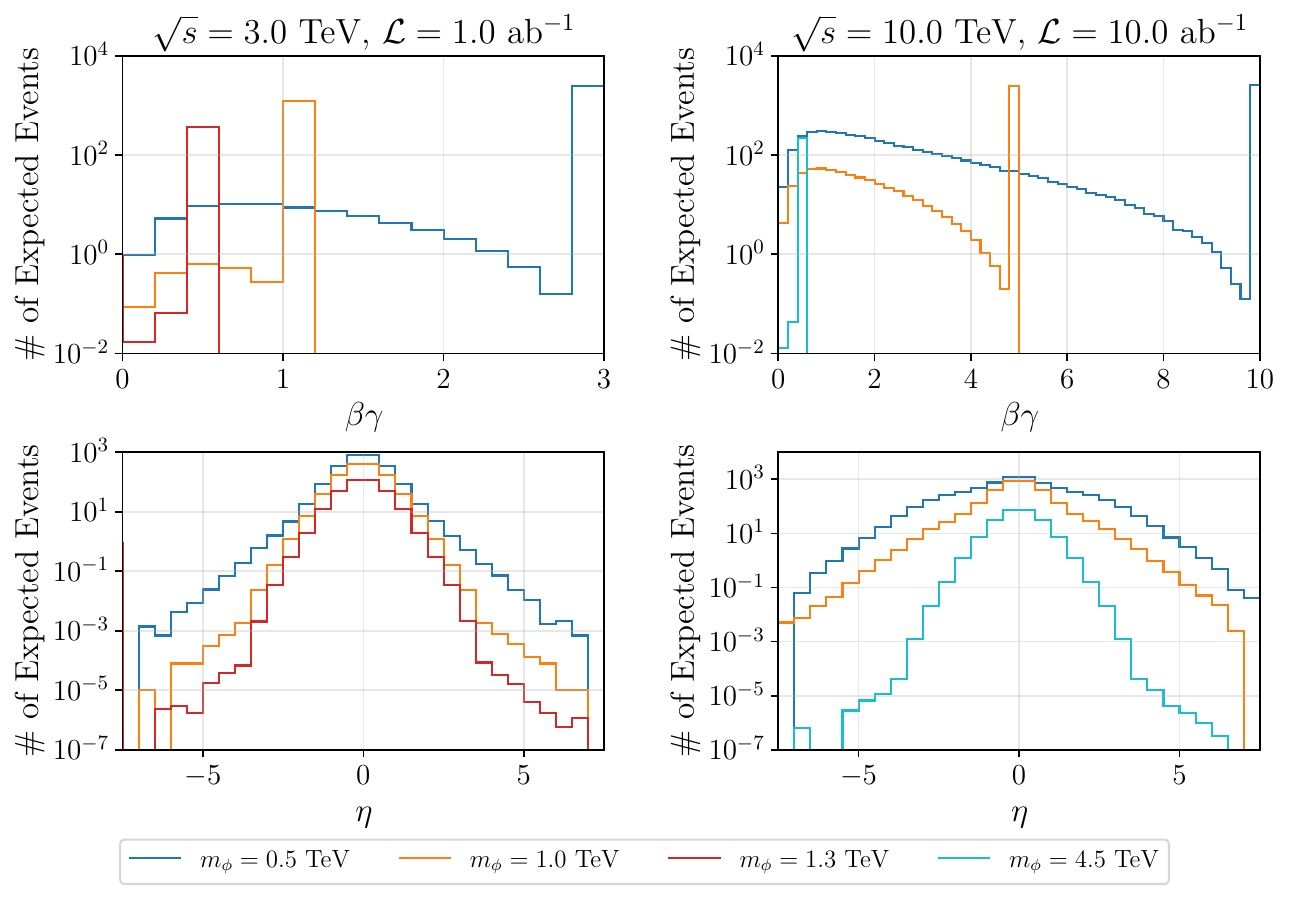}
    }
    \caption{Distribution of events in $\phi$'s $\beta \gamma$ (\textbf{top}) and $\eta$ (\textbf{bottom}) for different mediator masses at a MuC with center of mass energy of $\sqrt{s}=3$~TeV (\textbf{left}) or $\sqrt{s}=10$~TeV (\textbf{right}). The bin sizes are 0.2 and 0.5 for $\beta \gamma$ and $\eta$, respectively. The peak feature in the $\beta \gamma$ is due to events from the DY production channel, see Figure~\ref{fig:collider_diagrams}.}
    \label{fig:dist1Dbetaeta}
\end{figure}

We study the distribution of $\beta \gamma$, where $\beta=v/c$ and $\gamma=(1-\beta^2)^{-1/2}$, and pseudorapidity $\eta$.
In Figure~\ref{fig:dist1Dbetaeta} we show 1D histograms of $\beta \gamma$ and $\eta$ for a few different points on our parameter space and for center of mass energies of $\sqrt{s}=3$~TeV and $\sqrt{s}=10$~TeV. 
In making these histograms we run \texttt{MadGraph5} with $10^5$ events for each production channel (DY and VBF). 
We then weigh the number of events from each production channel by its corresponding cross-section to get the combined distribution. 

In Figure~\ref{fig:dist1Dbetaeta} we see features that are distinct to the individual production channels, DY and VBF, which in turn help inform the types of cuts we can use in a search for our model. 
In the DY channel, the initial muons each carry $E=\sqrt{s}/2$, which in turn gives rise to the maximum value of $\gamma=E/m_\phi$ for a given $m_\phi$ and leads to the very sharp peak in the $\beta \gamma$ distribution. An increase in $m_\phi$ results in a smaller $\gamma$, so the peak of the $\beta\gamma$ distribution will shift to lower values.\footnote{Currently the muon PDF in a muon beam is not implemented in \texttt{MadGraph5}. Thus, in our simulations we assumed each muon enters the interaction with $\sqrt{s}/2$ energy as an approximation. The PDF of a muon in the beam has support for $x \neq 1$ as well \cite{Han:2020uid,Han:2021kes,Ruiz:2021tdt}. As a result, the real distribution of $\beta \gamma$ from this channel is slightly diffused to values smaller than $\sqrt{s}/2$. Including the PDF will soften the peak in the distribution of events from the DY channel. Nonetheless, given the sharp peak in a muon PDF \cite{Han:2020uid,Han:2021kes,Ruiz:2021tdt}, we still expect a large peak in various kinematic distributions from this channel if the real muon PDF is included in the simulation. } 
We find that this channel gives rise to centralized $\phi$ particles (small $|\eta|$) as well. This motivates searches focusing on this region.
The VBF production channel gives rise to a wider spread in both $\beta \gamma$ and $\eta$, owing to the wider spread of initial gauge bosons PDF \cite{Han:2020uid,Han:2021kes,Ruiz:2021tdt}.

\subsection{Background and Time-of-flight}
\label{subsec:LLP_time_fly}

The background for LLPs can be divided into a reducible SM background, and an irreducible background from the detector response. 
While there are a handful of SM particles that appear as LLPs at a collider (\textit{e.g.} see Ref.~\cite{Alimena:2019zri}), using various kinematic observables, such as time-of-flight or anomalous ionization rate ($dE/dx$), allows us to distinguish them from heavy new physics LLPs. 
Other techniques such as empty bunch crossing also allows us to better understand and reject SM background; see Ref.~\cite{Lee:2018pag} for further discussions of SM background mitigation techniques.

\begin{figure}
    \centering
    \resizebox{\columnwidth}{!}{
    \includegraphics{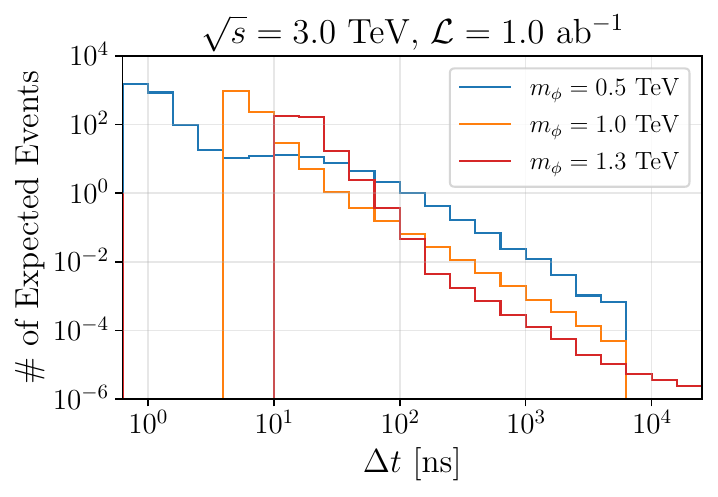}
    \includegraphics{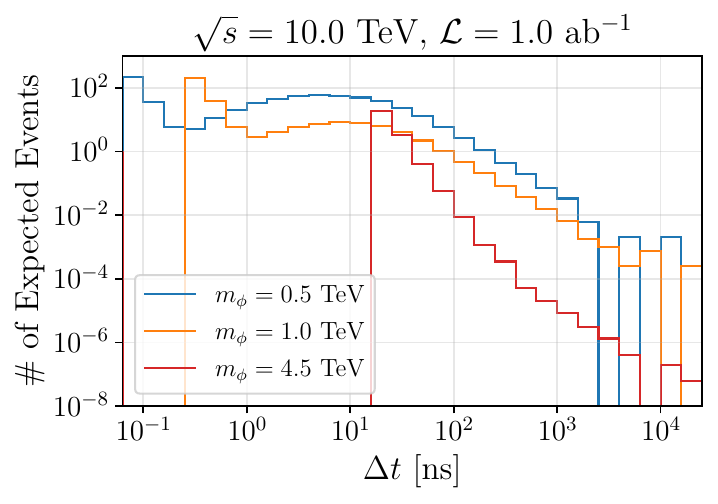}
    }
    \caption{Distribution of the difference between the arrival time of $\phi$ and a massless particle moving in the same direction at the onset of the muon system ($L=446.1$~cm), as predicted in our  \texttt{MadGraph5} simulation for three different mediator masses. In these histograms the log-scale x-axis is binned in steps of 0.2 decades. The arrival time scales linearly with $L$. Events at lower (higher) $\Delta t$ values correspond to the DY (VBF) production channel. As the mediator mass increases, the LLPs move more slowly, shifting the distributions to higher $\Delta t$ values.  }
    \label{fig:deltat}
\end{figure}

A useful quantity in reducing the SM background is the time-of-flight.\footnote{Another useful quantity in LLP searches is the anomalous ionization rate $dE/dx$. In particular, this quantity can be used in reconstructing the track mass and cutting away SM background. 
At LHC, measuring $dE/dx$ has a smaller uncertainty when only the barrel region pixel layers are used. 
This motivates LLP searches focusing on the central events at a MuC as well.
A proper calculation of this rate depends on characteristics of the detector material and tracks interactions with the detector, which can only be extracted from simulations. 
As a result, we acknowledge the possibility of using this quantity and its importance in the LLP searches, leaving more in-depth studies for future.}
In particular, the difference between the arrival time of a $\phi$ particle and SM particles in the same direction, $\Delta t=t_{\rm LLP}-t_{\rm SM}$, can be used to distinguish it from SM particles, which move with $\beta=1$. 
In Figure~\ref{fig:deltat} we show the distribution of this variable, at the inner border of the muon system ($L = 446.1$~cm transverse distance from the beam in the design of Table~\ref{tab:detector}). 
The time difference scales linearly with the transverse distance $L$ at which we measure it, thus we can use this figure to calculate the time difference at different transverse distances in the detector. 

Small $\Delta t$ events in Figure~\ref{fig:deltat} are produced by the DY initial channel (with larger $\beta$ and $\beta \gamma$ values), while events from the VBF channel arrive later since they have lower values of $\beta$. 
This figure shows that, at $L=446.1$~cm, timing resolution of around $0.1-1$~ns ($1-100$~ns) is enough to separate most events from the DY (VBF) channel from SM background. 
When compared to the current timing resolution at LHC ($\sim0.1-1$~ns for comparable $L$ values), we find that majority of events, especially from the VBF channel or for large enough $m_\phi$ values, arrive with enough delay that we can use the time-of-flight information to discern our LLP signal from background.

In addition to the SM reducible background discussed above, interactions of the beam, various tracks, or the BIB, with the material in the detector give rise to an irreducible background. 
Currently, there are limited studies on this topic \cite{Capdevilla:2021fmj}. Nonetheless, they have shown that such background events can be mitigated by timing information, track quality, and tracks directionality information. 
Further studies are in order to completely understand the detector response and background for LLP searches at a high energy MuC and, depending on the lifetime, different background mitigation strategies have to be deployed.

To be able to discover our model, we need to estimate this irreducible background as well. 
This can only be done with extensive simulations, beyond the scope of this work. 
Nonetheless, the signal yield of our model sets a reasonable target precision. Given the theoretical motivation of our model, and how generic its signal yield is for all models of electroweakly-generated LLPs, we hope our results in the upcoming section inform the on-going studies in developing the detector design.

\subsection{Charged Tracks and Displaced Leptons Signal}
\label{subsec:LLP_search}

In this section we calculate the signal yield of our model on the LLP part of the parameter space. We will show that our model gives rise to many events across the entire kinetically-accessible part of the parameter space. 
We will, in particular, highlight a double-peak feature in the distribution of events in the detector, and argue that it is a shared signature of all electroweakly-charged LLPs at a high energy MuC.

The kinematic quantities $\beta\gamma$ and $\eta$ directly enter the calculation of the trajectory of LLPs inside the detector. 
In LLP searches we are in particular interested in the transverse direction away from the beam, $L$. 
It can be shown that for realistic magnetic field values, the track curvature for $m_\phi \gtrsim 1$~TeV is very small and will not affect the LLP observables under study here.\footnote{For precise measurements of some LLP observables, such as closest approach of the track to the primary vertex, one needs to include the effect of the magnetic field. However, this is not the case for the displaced leptons and charged tracks signals studied in this work.} Thus, we neglect the effect of the magnetic field and assume the tracks move on straight lines after their production.

We can show that as a function of time, the transverse distance from the beam at time $t$ (in the lab frame) is given by 
\begin{equation}
    L(t, \beta, \eta)= \frac{\beta t }{\cosh \eta }  .
    \label{eq:L_T}
\end{equation}
Using this we can calculate the transverse displacement for every event at $t=\gamma \tau_\phi$.
The time $t=\gamma \tau_\phi$, is the average time at which the $\phi$ particle decays to a charged lepton and missing energy, giving rise to a displaced lepton.

\begin{figure}
    \centering
    \resizebox{\columnwidth}{!}{
    \includegraphics{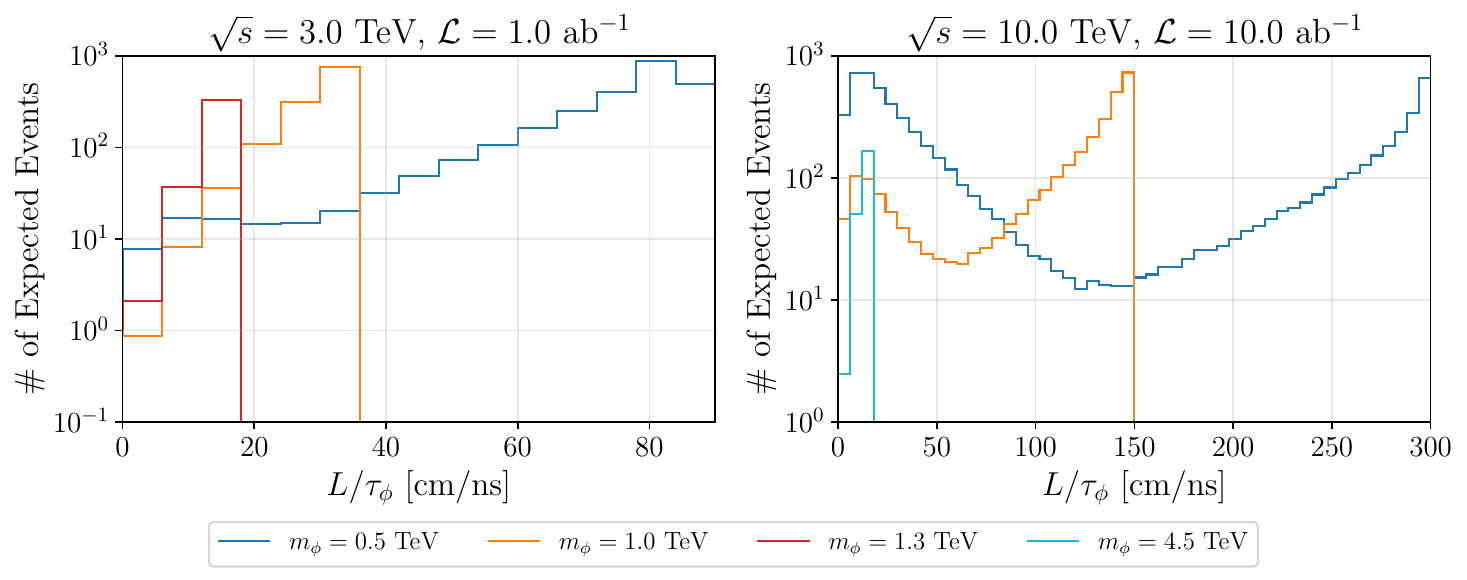}
    }
    \caption{Distribution of events in $\phi$'s transverse displacement at $t=\gamma \tau_\phi$ normalized by the lifetime, $L/\tau_\phi$ for different mediator masses at a MuC with center of mass energy of $\sqrt{s}=3$~TeV (\textbf{left}) or $\sqrt{s}=10$~TeV (\textbf{right}). The bin sizes are 6 cm/ns. The shape of the $L/\tau_\phi$ can be inferred from distributions in Figure~\ref{fig:dist1Dbetaeta} and Eq.~\eqref{eq:L_T}. For different points in the parameter space the mediator lifetime $\tau_\phi$ can be read off of Figure~\ref{fig:lifetime}; combined with these plots, we can then find the transverse displacement for events at any point in the parameter space for every DM mass. The double-peak feature in the right figure can give rise to interesting LLP signals, as elaborated in the text. }
    \label{fig:dist1DLtau}
\end{figure}

In Figure~\ref{fig:dist1DLtau} we show the distribution of transverse displacement $L$ at $t=\gamma \tau_\phi$ for a few different points in our parameter space. 
The sharp peak in large $L/\tau_\phi$ values in the distributions corresponds to events from the DY  channel that all have a very large $\beta \gamma$ value. We also find that for smaller mediator masses the distribution stretches to larger values of $L/\tau_\phi$, giving rise to more displaced leptons further away from the beam.

The $L/\tau_\phi$ distribution exhibits an interesting double-peak feature; the peak at large values arises from the DY production while the peak at lower values is an artifact of the $\eta$ dependence in Eq.~\eqref{eq:L_T} and arise from VBF channel. 
Since these production channels universally apply to all electroweakly-charged LLPs, we expect a similar double-peak feature in the distribution of $L/\tau_\phi$ in all such models.\footnote{Once the muon PDF is properly included in the simulations, we expect this peak to spread out to some extent. Nonetheless, the sharp peak at $x \rightarrow 1$ in this PDF \cite{Han:2020uid,Han:2021kes,Ruiz:2021tdt} still will give rise to a peak in the $\beta \gamma$ and in $L/\tau_\phi$ distributions, maintaining this double-peak feature. } 
The main irreducible background for LLP searches come from secondary interactions of particles with the detector material and this double-peak feature may not be manifested in this background. As a result, this feature could be a smoking gun signature of electroweakly-charged LLPs at a MuC and could be used for an efficient background rejection.

\begin{figure}
    \centering
    \resizebox{\columnwidth}{!}{
    \includegraphics{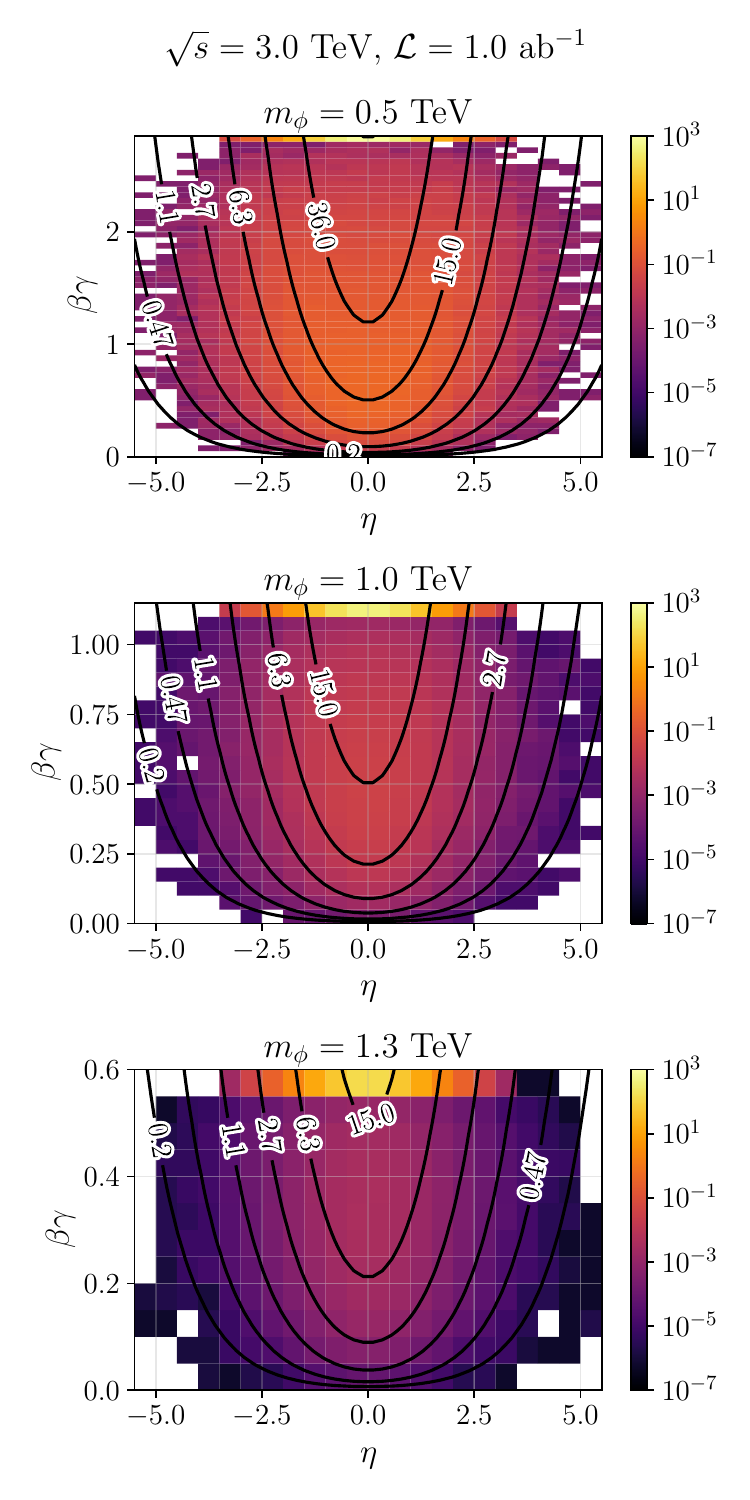}
    \includegraphics{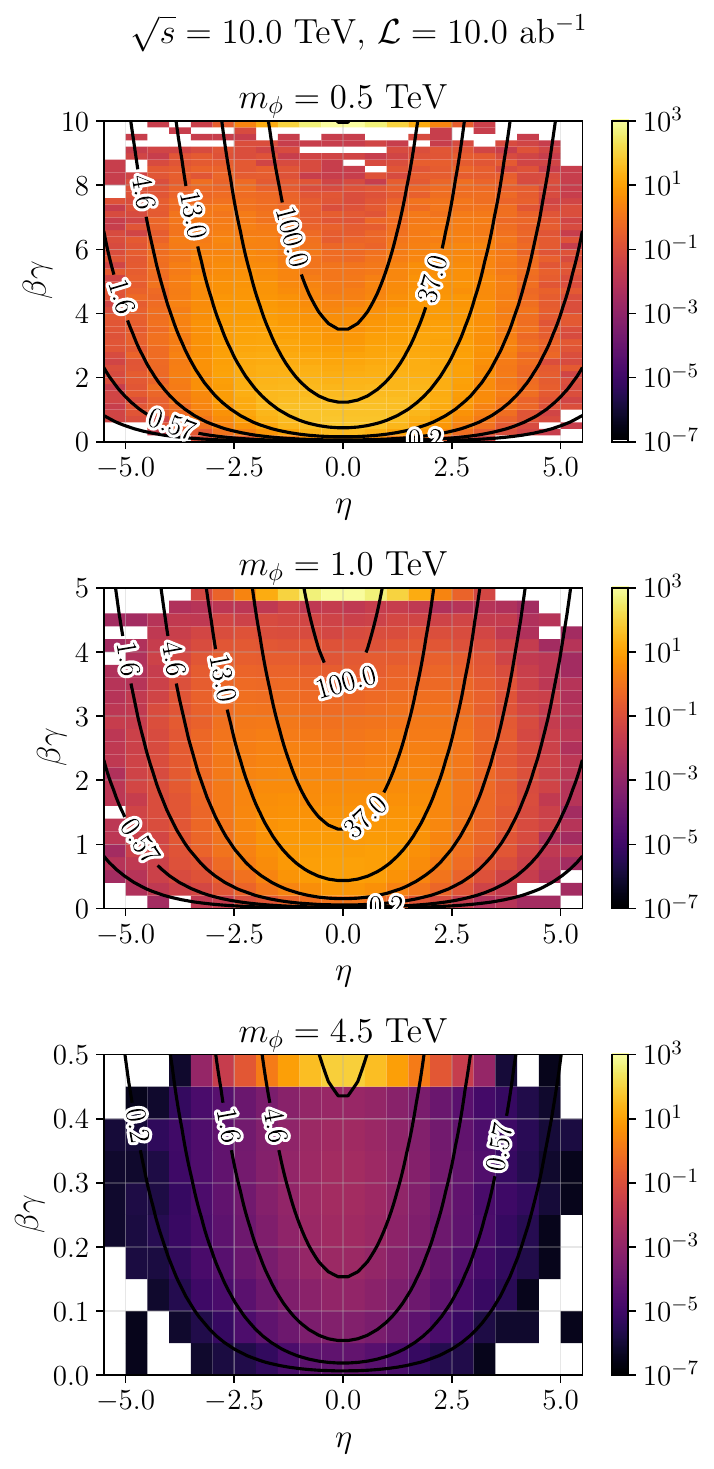}
    }
    \caption{Distribution of events in the $\eta-\beta \gamma$ plane for different $\phi$ masses at a 3 (10) TeV MuC on \textbf{left} (\textbf{right}). The bin sizes are 0.5 for $\eta$. For $\beta \gamma$, the bin size is 0.05 except for $\sqrt{s} = 10$ TeV and $m_\phi \leq 1$ TeV, where we use 0.2. Contours of constant $L/\tau_\phi$ [cm/ns] are shown as well (see Eq.~\eqref{eq:L_T}), see the text for further explanation on the shape of the contours. The cluster of events at maximum $\beta\gamma$ values are from the DY production channel. We find that majority of events from the DY and the VBF channel appear with different kinematics. }
    \label{fig:dist2Dbetaeta}
\end{figure}

In Figure~\ref{fig:dist2Dbetaeta} we show the joint distribution of events on the plane of $\eta-\beta \gamma$ for a few different mediator masses. 
In each plot we can see a cluster of events at highest physically possible value of $\beta \gamma$ which correspond to the DY-generated events. The VBF-generated events have a wider distribution in both $\eta$ and $\beta \gamma$, as expected.

We also show contours of constant $L/\tau_\phi$ on the $\beta \gamma - \eta$ plane in Figure~\ref{fig:dist2Dbetaeta}. 
The shape of these contours can be inferred from Eq.~\eqref{eq:L_T}. 
For a fixed value of $\tau_\phi$, these contours tell us at what transverse distances the $\phi$ particle will (on average) decay and inform the signal yield of our model.  
For $\tau_\phi=1$~ns these contours show $L$ in cm, while for larger lifetimes, the contours of constant $L$ move down on the plots. 
We can repeat this calculation for all mediator and DM masses to calculate the rate for displaced leptons appearing in any transverse distance segment of the detector, inclusive over other kinematic variables such as $\eta$ and $\beta \gamma$. 
The rate for displaced leptons appearing at a given transverse distance is our model's main signature in the LLP region.

\begin{figure}
    \centering
    \resizebox{\columnwidth}{!}{
    \includegraphics{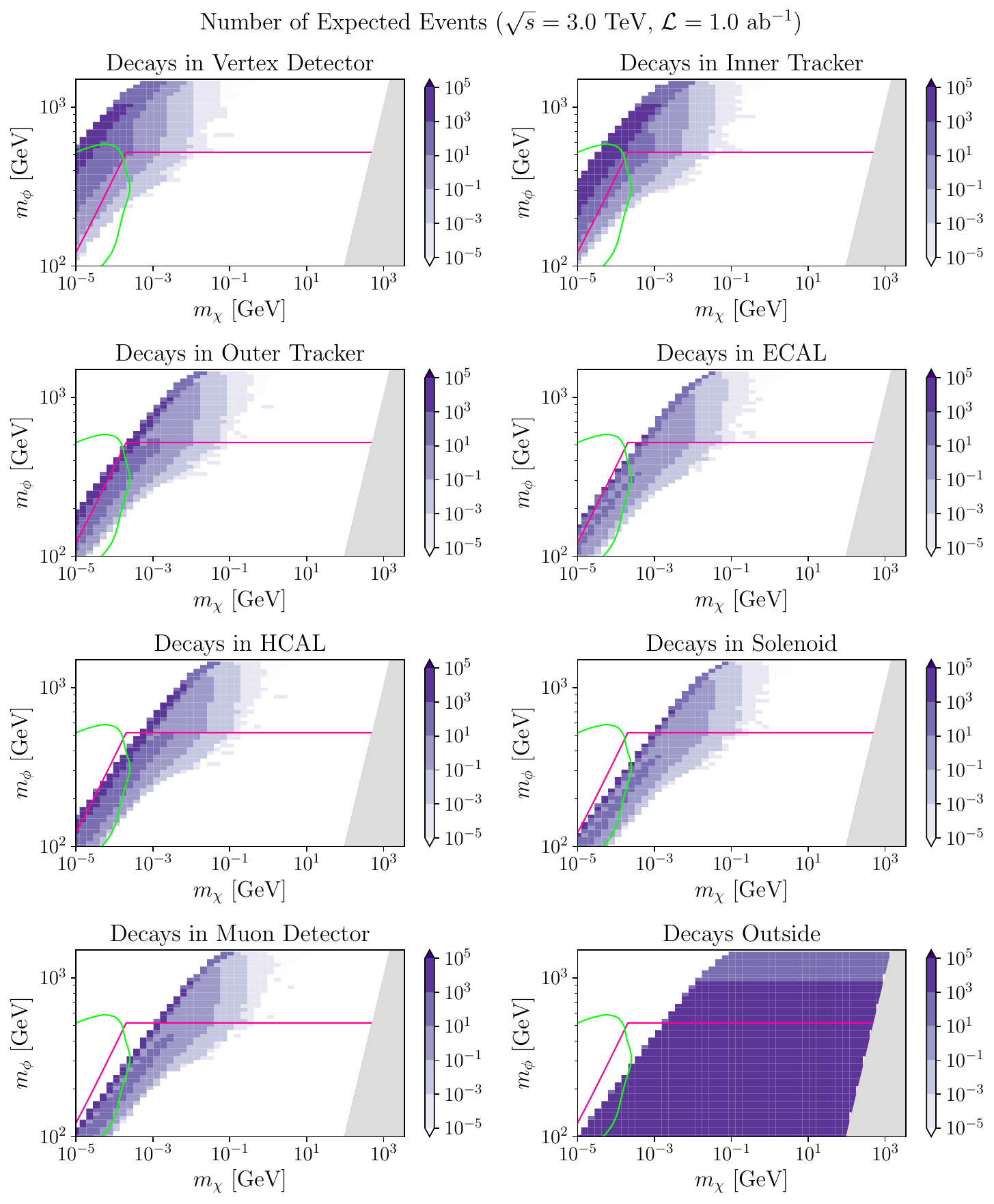}
    }
    \caption{Average rate of displaced leptons in different barrel regions of the detector for a 3 TeV MuC, assuming the tentative design of Table~\ref{tab:detector}, as well as average number of detector-stable charged tracks. The DM Yukawa coupling to the mediator is chosen to get the right relic abundance today. The left (right) cluster corresponds to events from the initial DY (VBF) channel of Figure~\ref{fig:collider_diagrams}, see the text for further explanation about the shape of contours. The gray region corresponds to $m_\chi > m_\phi$ and is not phenomenologically viable. Our model's signal yield can serve as a benchmark in studies of tolerable systematics in LLP searches at a MuC. The region below the green (pink) line is already ruled out by the LHC search in Ref.~\cite{ATLAS:2020wjh} (Ref.~\cite{CMS:2024qys}). }
    \label{fig:money_LLP3}
\end{figure}

\begin{figure}
    \centering
    \resizebox{\columnwidth}{!}{
    \includegraphics{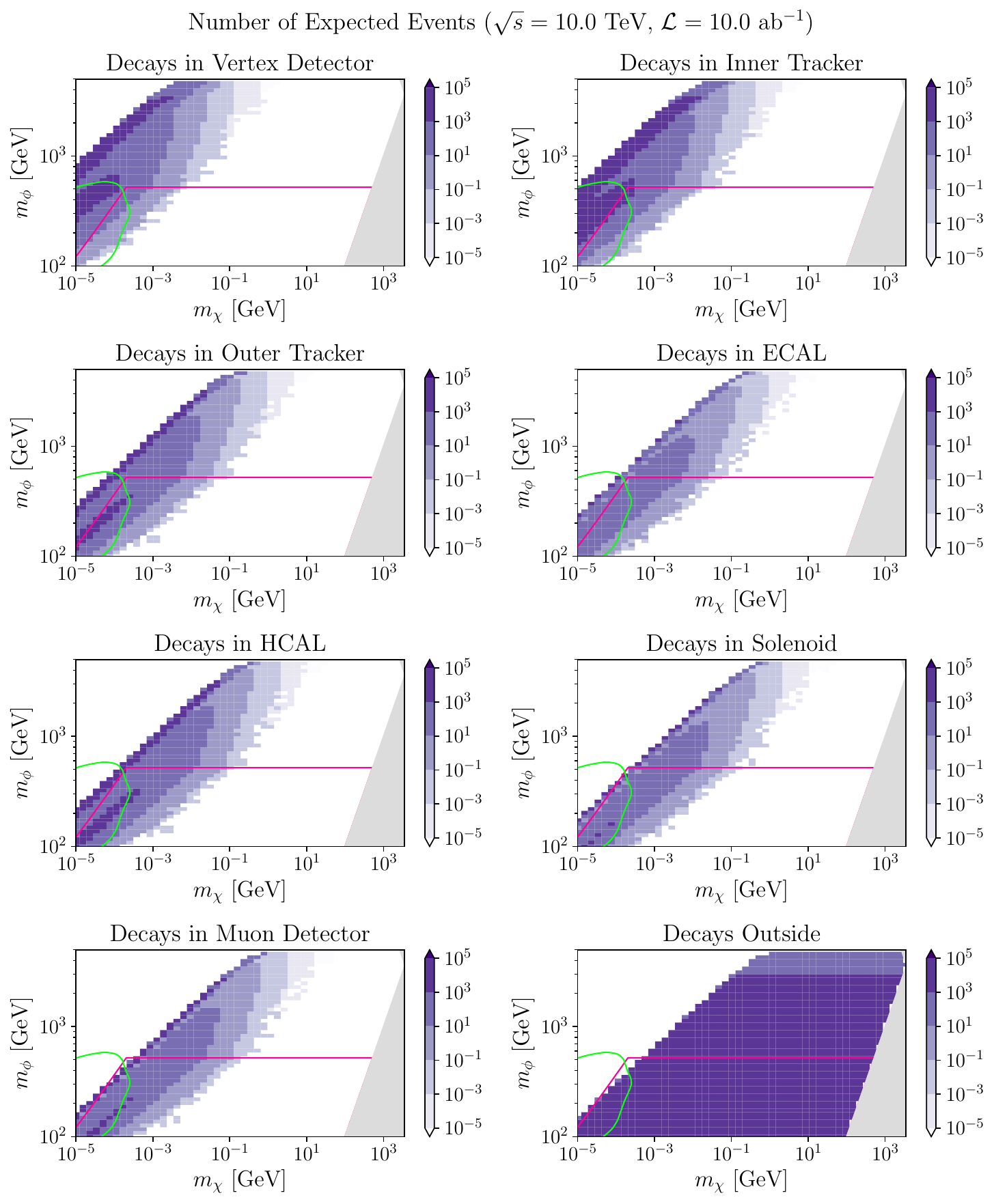}
    }
    \caption{Similar to Figure~\ref{fig:money_LLP3}, but for a MuC with center of mass energy of $\sqrt{s}=10$ TeV. We again find that virtually the entire kinematically-accessible part of the parameter space gives rise to large numbers of charged tracks and displaced leptons. The region below the green (pink) line is already ruled out by the LHC search in Ref.~\cite{ATLAS:2020wjh} (Ref.~\cite{CMS:2024qys}). }
    \label{fig:money_LLP10}
\end{figure}

\begin{figure}
    \centering
    \resizebox{\columnwidth}{!}{    \includegraphics{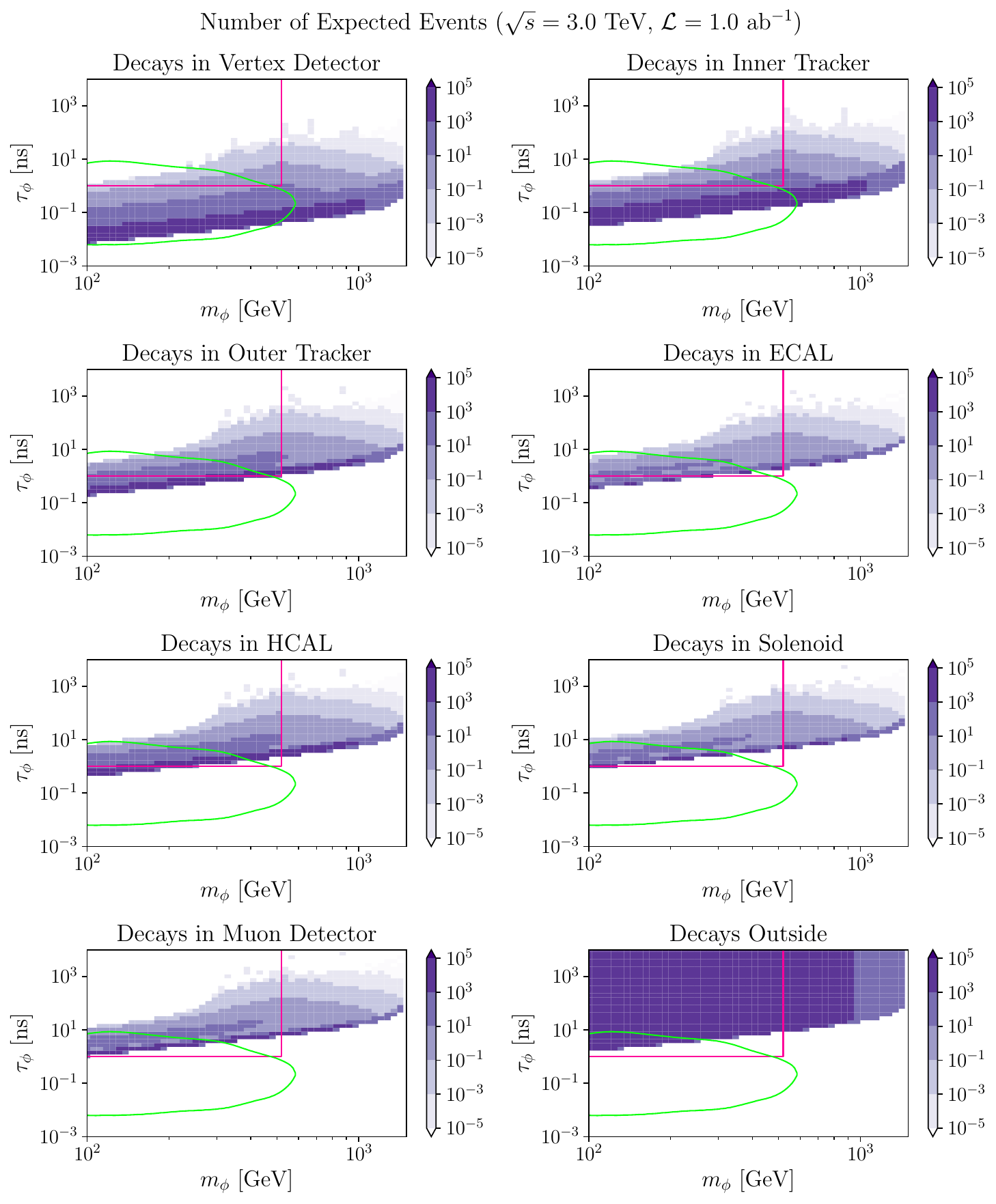}
    }
    \caption{Same as Figure~\ref{fig:money_LLP3} but on the plane of $m_\phi$-$\tau_\phi$. We find a large signal yield, in at least one component of the detector, for most of the parameter space beyond the current LHC reach. The region below the green (pink) line is already ruled out by the LHC search in Ref.~\cite{ATLAS:2020wjh} (Ref.~\cite{CMS:2024qys}). }
    \label{fig:money_tau_LLP3}
\end{figure}

\begin{figure}
    \centering
    \resizebox{\columnwidth}{!}{ \includegraphics{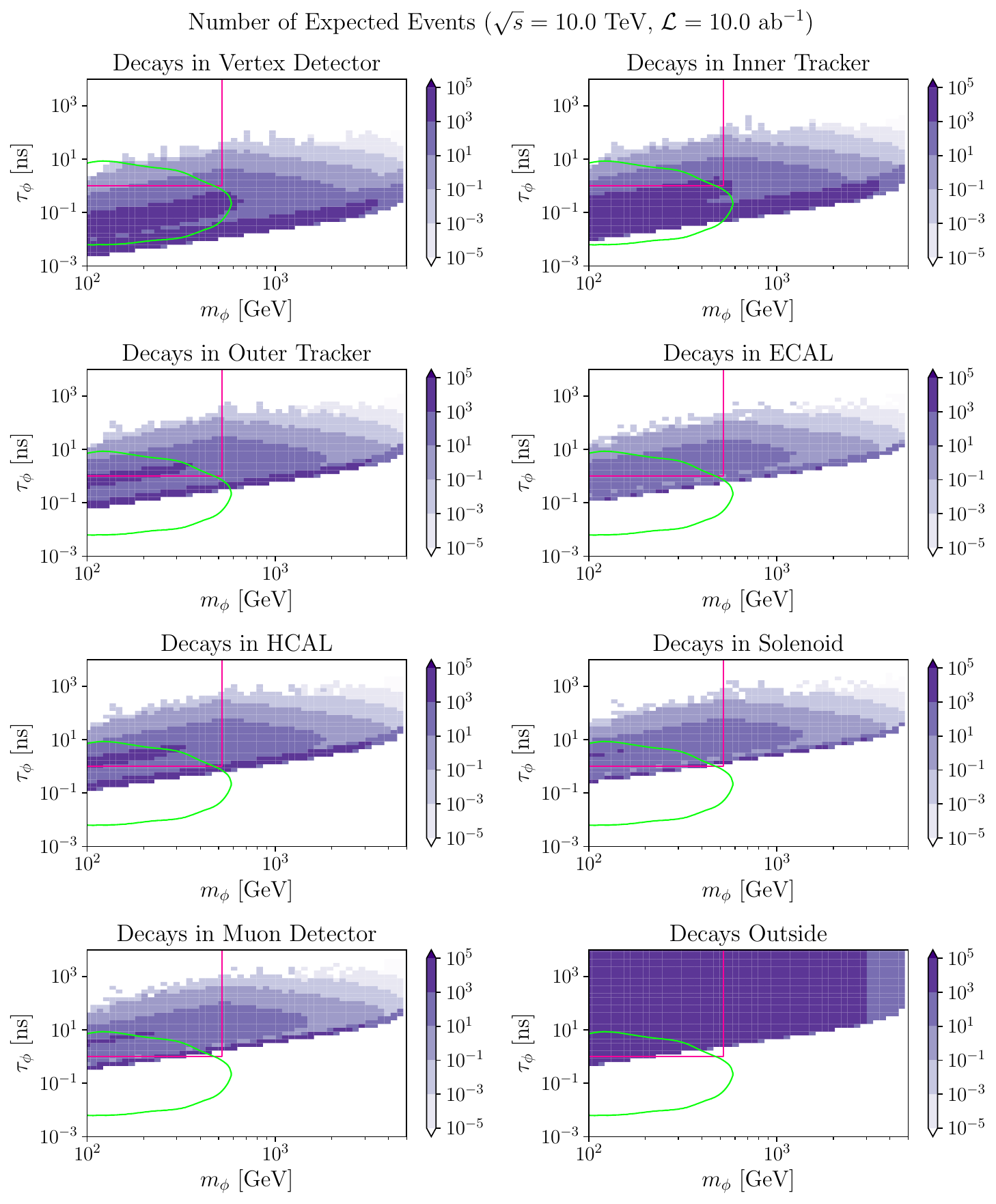}
    }
    \caption{Same as Figure~\ref{fig:money_LLP10} but on the plane of $m_\phi$-$\tau_\phi$. We find a large signal yield, in at least one component of the detector, for most of the parameter space beyond the current LHC reach. The region below the green (pink) line is already ruled out by the LHC search in Ref.~\cite{ATLAS:2020wjh} (Ref.~\cite{CMS:2024qys}). }
    \label{fig:money_tau_LLP10}
\end{figure}

The transverse distance can be mapped to detector components as defined in Table~\ref{tab:detector}.  
We show the average number of displaced leptons in the barrel region of each detector component on our model parameter space in Figures~\ref{fig:money_LLP3} and \ref{fig:money_tau_LLP3} (Figures~\ref{fig:money_LLP10} and \ref{fig:money_tau_LLP10}) for $\sqrt{s}=3$~TeV ($\sqrt{s}=10$~TeV).\footnote{In the compressed region, $m_\phi \approx m_\chi$, the daughter lepton from $\phi$ decay will be soft and will typically escape detection, giving rise to a disappearing track if they decay inside the detector. 
However, the lifetime $\tau_\phi$ is very long in this region of the parameter space (see Figure~\ref{fig:lifetime}), so our signal will still be a heavy stable charged track.
The disappearing track signals for LLPs that decay inside a MuC are studied extensively in Ref.~\cite{Capdevilla:2021fmj}.} 
Endcap regions are neglected since, similar to LHC, it is reasonable to expect more SM background in endcaps with large track masses that are difficult to distinguish from LLP signals, and that $dE/dx$ measurements in the endcaps are slightly less accurate compared to their barrel region counterparts. 
Our analysis can be straightforwardly repeated for the endcap regions as well.

In Figures~\ref{fig:money_LLP3} and \ref{fig:money_LLP10} we show the signal yield on the plane of $m_\chi$-$m_\phi$. 
In each detector component, the distribution of events clearly divides up into two separate regions. The  upper cluster of events in these figures - at lower $m_\chi$ and higher $m_\phi$ values - corresponds to events originating from initial DY channel, while the other cluster of events come from the initial VBF channel. 
These two separate clusters of events are a manifestation of the previously explained ``double-peak" feature seen in Figure~\ref{fig:dist1DLtau}.

The features of the signal regions in Figures~\ref{fig:money_LLP3} and \ref{fig:money_LLP10} can be understood through the shape of the $L$ contours in Figure~\ref{fig:dist2Dbetaeta}. 
Recall that for a fixed value of $\tau_\phi$, the contours in Figure~\ref{fig:dist2Dbetaeta} tell us at what transverse distances the $\phi$ particle will (on average) decay, $i.e.$ $L(t=\gamma \tau_\phi)$. 
At very short lifetimes, the contours associated with each barrel region in Figure~\ref{fig:dist2Dbetaeta} are above the entire distribution of events. 
As the lifetime increases, contours of constant $L$ move down on Figure~\ref{fig:dist2Dbetaeta} (see Figure~\ref{fig:lifetime} for contours of constant lifetime). 
For fixed $m_\phi$, $\tau_\phi$ increases with higher $m_\chi$ values, and so we find that the events at low $m_\chi$ are predominantly from the DY channel (as they are on top of 2D histograms in Figure~\ref{fig:dist2Dbetaeta}). As we increase $m_\chi$ (and $\tau_\phi$), the VBF channel increasingly contributes. 
Since distributions of DY- and VBF-generated events peak at different $\beta \gamma$ values in Figure~\ref{fig:dist2Dbetaeta}, we find two separate cluster of events in each detector component in Figures~\ref{fig:money_LLP3} and \ref{fig:money_LLP10}. 
The slope of each cluster of events is inherited from contours of constant lifetime on the mass plane, see Figure~\ref{fig:lifetime}. 

The last panel in both Figures~\ref{fig:money_LLP3} and \ref{fig:money_LLP10} corresponds to tracks that are expected to go outside the detectors without decaying. 
We only include the tracks with small enough $\eta$s that go through all barrel regions, \textit{i.e.} $\eta \lesssim 0.61$, see Table~\ref{tab:detector}. 
These panels clearly show a large number of such stable tracks are expected on a large part of the viable parameter space, allowing us to look for the signals of this model on virtually the entire kinematically accessible parameter space.

In Figures~\ref{fig:money_tau_LLP3} and \ref{fig:money_tau_LLP10} we present the same information (the signal yield in each detector component) on the $m_\phi$-$\tau_\phi$ plane. 
Existing LHC bounds on the parameter space are marked again. We clearly see that at a MuC, a large part of the parameter space unaccessible at LHC will have a large signal yield. 
The double-peak feature discussed above is clearly visible in these plots as well.

Our results should be contrasted to the current bounds from LHC on Figure~\ref{fig:lifetime}. 
Putting the numbers from all panels of Figures~\ref{fig:money_LLP3}--\ref{fig:money_tau_LLP10} together, we find that we have at least $10^3$ events in at least one part of the detector for almost the entire kinematically-accessible parts of our parameter space, see Appendix~\ref{app:LLP}. 
Hence, once a reasonable degree of systematics is achieved, future high energy MuCs can sift through a much larger part of the parameter space than LHC. 
This clearly underscores the supremacy of a high energy MuC in searching for this DM model. 
Crucial in this supremacy are the stable tracks that go through the entire detector without decaying, see the last panel of Figures~\ref{fig:money_LLP3}--\ref{fig:money_tau_LLP10}. For most of the parameter space this is the dominant LLP signal and can be used for probing the parameter space inaccessible by displaced lepton signals in the detector. 
The double-peak feature, which in turn is an artifact of two distinct production channels, suggests for many points in the parameter space we can have many displaced leptons in more than one component of the detector.

We should emphasize that Figures~\ref{fig:money_LLP3}--\ref{fig:money_tau_LLP10} show the {\it average} number of decays and displaced leptons (or stable charged tracks). The true number of events in each barrel region will be drawn from a Poisson distribution around this average. 
Further details about kinematics of our model in the LLP region are included in Appendix~\ref{app:LLP}.

\section{Conclusions}
\label{sec:conclusion}

We initiated a detailed study of the freeze-in flavored DM setup in which flavored DM coupling to the slepton-like mediator is very small. We showed that 
the interplay of direct freeze-in and the mediator freeze-out gives rise to an interesting abundance calculation, such that the dark matter mass should be below a few TeV in order to avoid overclosing the universe. 
Combined with the existing collider and astrophysical bounds, this gives rise to a bounded viable parameter space for this model.

We also studied signals of our model at a future high energy Muon Collider (MuC). 
As a result of the feeble Yukawa coupling of DM, all signals of the model in a MuC originate from the on-shell pair production of the mediator. 
The range of mediator masses and lifetimes in the viable parameter space is such that we can have prompt, as well as various long-lived particle signals at a MuC. 
We divided our parameter space into two broad ranges according to the mediator's lifetime, namely prompt and LLP regions.

In the prompt decay region, we proposed a rudimentary analysis and identified the target systematics that ought to be achieved in order to discover our model at a MuC. 
Our analysis uses cuts on invariant leptonic system mass $m_{\ell\ell}$, missing transverse mass parameter $M_{T2}$, and the leptonic system transverse momentum $p_{T,\ell\ell}$.
We found that in the part of the parameter space with prompt mediator decay, if systematic uncertainties are comparable or smaller than SM background, our proposed analysis can discover this model at effectively the entire kinematically-accessible parameter space.

For longer lifetimes, we will have a charged track in the detector which, depending on the lifetime, can either decay to a lepton and missing energy or decay outside the detector. 
While the SM background can be reduced away using quantities such as time-of-flight or anomalous ionization rate, detector response can give rise to an irreducible background. 
A proper study of these effects requires extensive simulations and is left for future works. In the absence of concrete simulations, here we reported the signal yield and various kinematics distributions. 
In particular, we showed that the displaced lepton distribution has a double-peak signature that could potentially be used for efficient background rejection. This feature appears as two large sets of signal events, produced through the two distinct channels (DY and VBF), in different detector components.

Our work can be extended in many directions. Foremost among them is a detailed study of the detector design that enables detailed studies of the background. 
Given the great reach of future MuCs in our model's parameter space, the universality of its production rate, and its theoretical motivation, we believe our model can serve as a motivated benchmark target for future studies on development of the detector by setting a target for systematics of such works.

Our analysis can also be repeated for other flavored DM models, \textit{i.e.} mediators with same charges as other SM fermions. 
All these setups have the added bonus that the artificial alignment of the DM Yukawa and SM Yukawas in flavored DM models, required for evading bounds from flavor-changing neutral currents and lepton flavor violation, are completely avoided thanks to the feeble DM coupling. 
While the relic abundance calculation resembles the calculation here, we expect a different collider phenomenology for some of them thanks to their different production and decay channels. 

Furthermore, as alluded to earlier, the muon PDF is not yet included in \texttt{MadGraph5}, thus we neglected its effect in the DY production channel. 
With the ever-growing interest in BSM signals at a MuC, it is interesting to work on proper embedding of this PDF in the event-generation pipeline.

It will also be interesting to consider signals of such models in other experiments, such as various astrophysical or direct detection searches. 
(See Refs.~\cite{Garny:2018ali,Decant:2021mhj} for a study of astrophysical bounds on a quark-flavored DM model in the freeze-in regime.)
 In particular, there is a large part of the parameter space inaccessible to colliders in foreseeable future, motivating searches in the aforementioned complementary fronts.

High energy MuCs are strongly motivated for their ability in probing different models of the Higgs boson UV-completion, flavorful new physics, and DM models. Our study outlines a simple and interesting DM model with unique signatures that can serve as a target for searches at a high energy MuC and inform its detector design.

\section*{Acknowledgement}

We thank Rodolfo Capdevilla, Spencer Chang, Jochen Heinrich, Samuel Homiller, Graham Kribs, and Ben Lillard for helpful discussions. We are especially grateful for numerous illuminating discussions with Laura Jeanty about LLPs. We also thank Samuel Homiller and our anonymous referee for constructive comments on the draft. The work of P.A. is supported by the U.S. Department of Energy under grant number DE-SC0011640. A.R. and T.-T.Y. are supported in part by NSF CAREER grant PHY-1944826.

The code to produce the plots and results in this paper is available \href{https://github.com/ariaradick/LFDM_at_MuC}{here}. We also use the following software:

\noindent \texttt{Julia} \cite{bezanson_julia_2017}, \texttt{DataFrames.jl} \cite{JSSv107i04}, \texttt{CSV.jl} \cite{jacob_quinn_2023_8004128}, \texttt{DifferentialEquations.jl} \cite{DifferentialEquations.jl-2017}, \texttt{HCubature.jl} \cite{HCubature}, \texttt{Optim.jl} \cite{Optim.jl-2018}, \texttt{PythonCall.jl} \cite{PythonCall.jl}, \texttt{QuadGK.jl} \cite{quadgk}, \texttt{Roots.jl} \cite{Roots.jl}, \texttt{matplotlib} \cite{Hunter:2007}.


\appendix

\section{More Details on Relic Abundance Calculation}
\label{app:abundance}

The relic abundance is directly related to the final $\chi$ yield, $Y_\chi^\infty$, 
\begin{align}
    \Omega_\chi &= 2 m_\chi Y_\chi^\infty \frac{T_0^3}{30} \frac{8 \pi G}{3 H_0^2}
\end{align}
where $T_0 = 2.7255$ K and $H_0 = 100 h$ km s$^{-1}$ Mpc$^{-1}$ are the temperature and Hubble constant today, respectively, and $G$ is the gravitational constant. The factor of $2$ accounts for $\chi$ and $\bar{\chi}$ both contributing to the final relic abundance. To be fully accurate, one should run the Boltzmann equation solver out to a sufficiently long time to obtain $Y_\chi^\infty$. However, we can also estimate this by noticing that $Y_\phi$ follows its freeze-out value until decaying, and $Y_\chi$ follows its freeze-in value (assuming $Y_\phi$ is in equilibrium) until $\phi$ decays. This information leads to the approximation that
\begin{align}
    Y_\chi^\infty \approx \ycfi + \ypfo
\end{align}
This approximation is useful because $\ypfo$ only depends on $\phi$'s mass, its couplings to the SM are fixed by its charge, and we can analytically calculate $\ycfi$. 
We have checked that for the majority of the parameter space, this approximation has less than $1\%$ error in the abundance calculation. 

Starting with $\chi$'s Boltzmann equation we set $Y_\phi = Y_{\phi, \textrm{EQ}}$ and start from $Y_\chi = 0$. The Boltzmann equation for $\chi$ will be \cite{Hall:2009bx}
\begin{align}
  \frac{d Y_\chi}{d x} &= \frac{x^3}{H(m_\phi)} \frac{g_\phi 3 \Gamma_{\phi \to \ell \chi}}{2 \pi^2}K_1(x),
\end{align}
and we can simply integrate this from $x = 0$ to $x = \infty$ to find
\begin{align}
  \ycfi = \frac{9 g_\phi \Gamma_{\phi \to \ell \chi}}{4 \pi H(m_\phi)},
\end{align}
with the decay width for $\phi \to \ell \chi$ given by
\begin{align}
    \Gamma_{\phi \to \ell \chi} &= \frac{\lambda^2 m_\phi}{16 \pi} \left( 1 - \frac{m_\chi^2}{m_\phi^2} \right)^2.
\end{align}
We can plug in the approximation to the relic abundance to get
\begin{align}
    \Omega_\chi &= 2 m_\chi \left( Y_\chi^{\textrm{FI}} + Y_\phi^{\textrm{FO}} \right) \frac{T_0^3}{30} \frac{8 \pi G}{3 H_0^2}
\end{align}
To find the particular $\lambda$ that gives the correct relic abundance $\Omega_\chi h^2 \approx 0.12$, we insert our expression for $Y_{\chi}^{\textrm{FI}}$ and rearrange to get the Yukawa coupling that gives rise to the right relic abundance today
\begin{align}
  \lambda' &= \frac{8 \pi}{3 (1 - m_\chi^2 / m_\phi^2)} \sqrt{\frac{H(m_\phi)}{g_\phi m_\chi m_\phi }} \sqrt{ \frac{45 \Omega_\chi H_0^2}{8 \pi G T_0^3} - m_\chi \ypfo }.
\end{align}
We plot this quantity in Figure~\ref{fig:lambda_parameter}.

\section{More Details on the Prompt Search}
\label{app:prompt}

In this appendix we provide further results of our \texttt{MadGraph5} simulations that inform the cuts we used in our prompt region search.

\begin{figure}
    \centering
    \resizebox{\columnwidth}{!}{
    \includegraphics{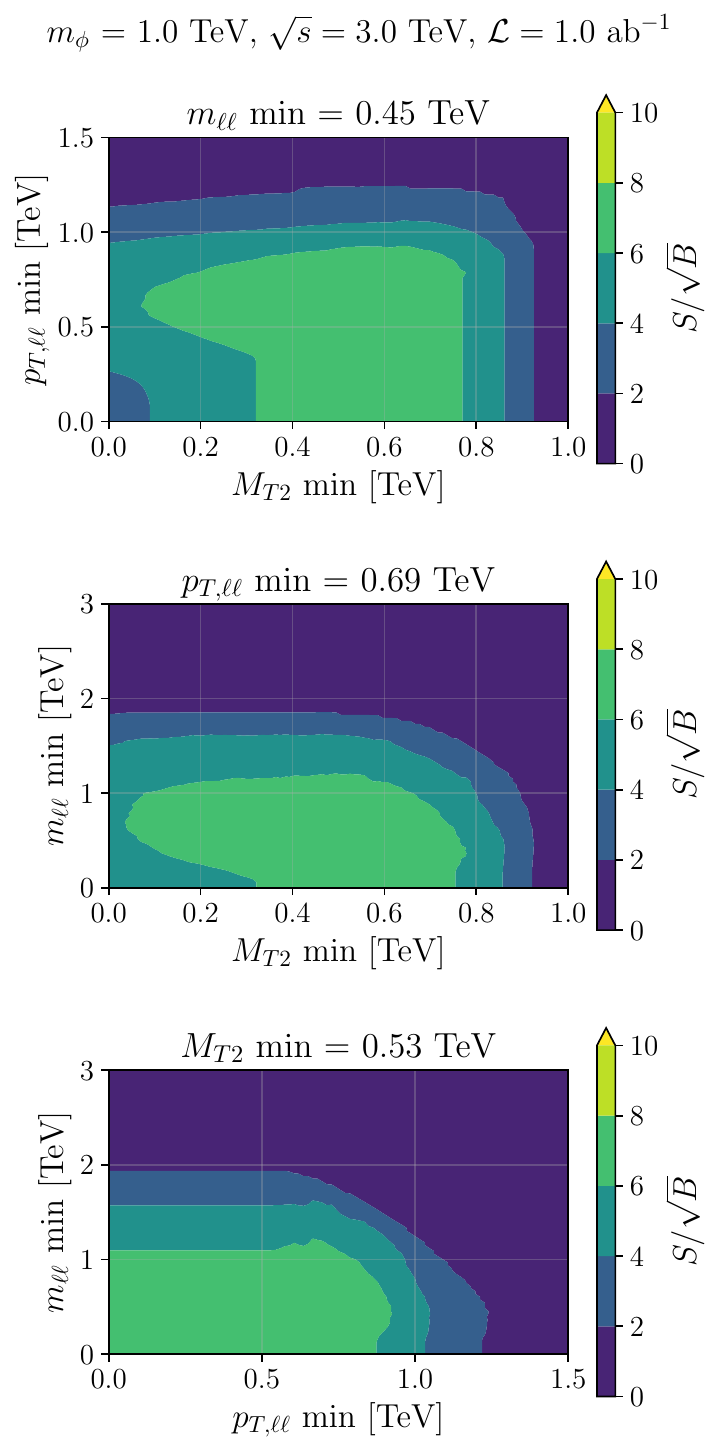}
    \includegraphics{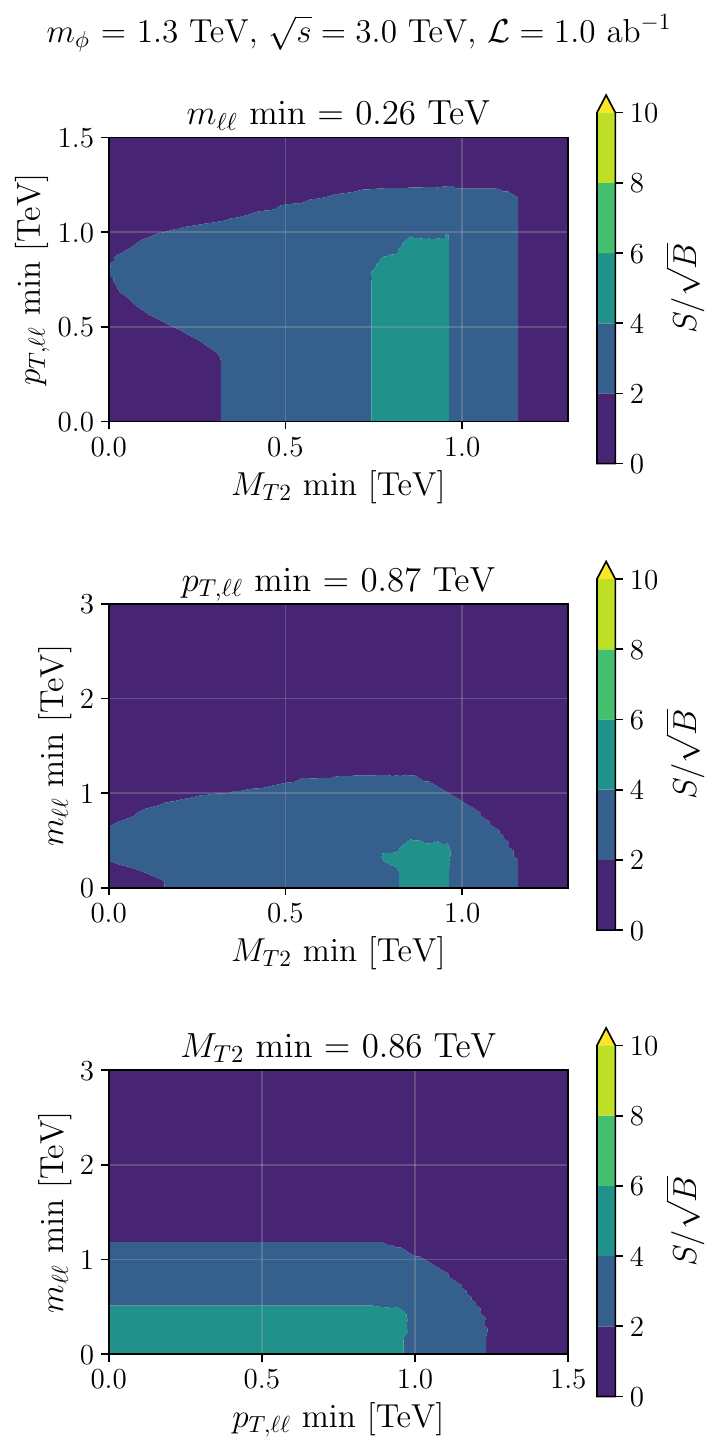}
    }
    \caption{Contours of significance of the signal (signal count $S$ over square root of background $\sqrt{B}$) for different minimum values on each observable. We show the effect of cuts on events from two different $\phi$ masses, 1 TeV (\textbf{left}) and 1.3 TeV (\textbf{right}), at a MuC with center of mass energy of $\sqrt{s}=3$~TeV. In each panel we show the effect of varying two cuts, while fixing the third one to the value that maximizes the significance. We find large significances are attainable. These figures inform the cuts used in definition of the signal regions.}
    \label{fig:histograms_2D_prompt_3}
\end{figure}

\begin{figure}
    \centering
    \resizebox{\columnwidth}{!}{
    \includegraphics{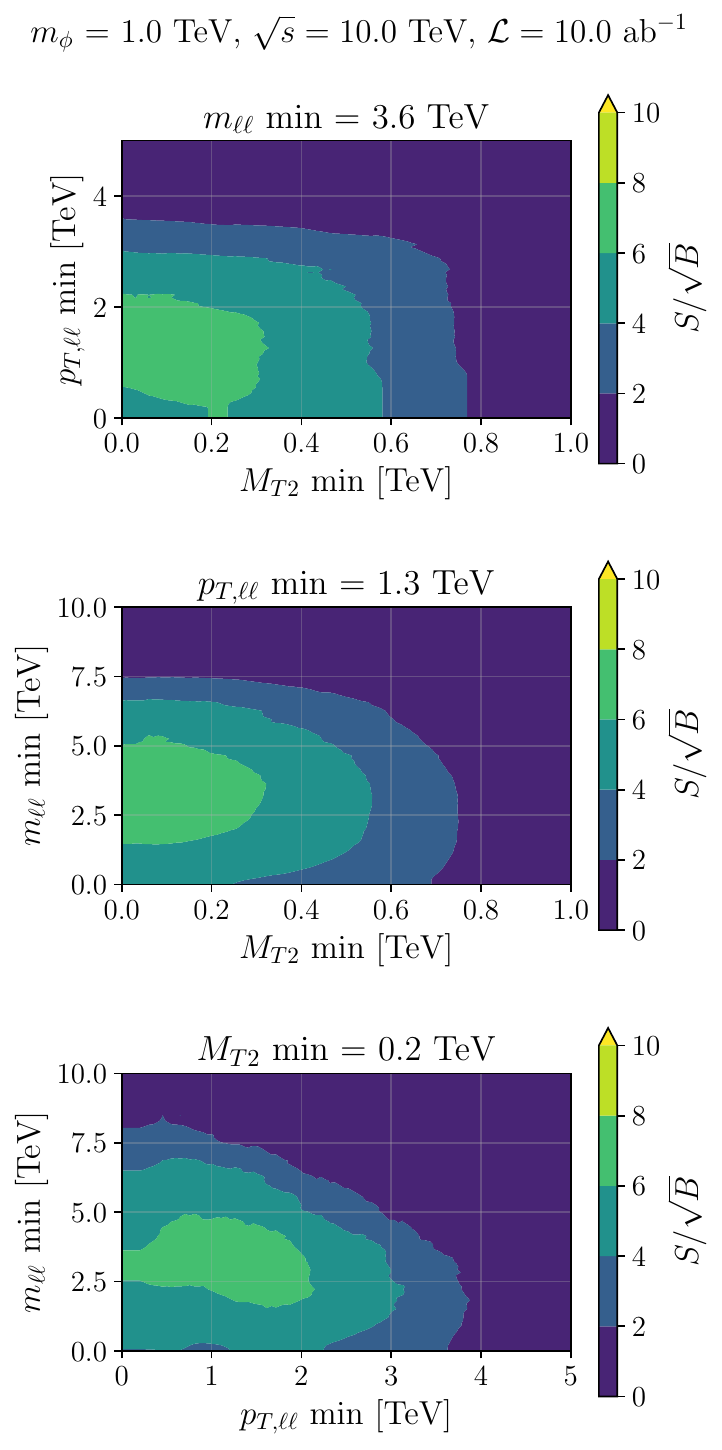}
    \includegraphics{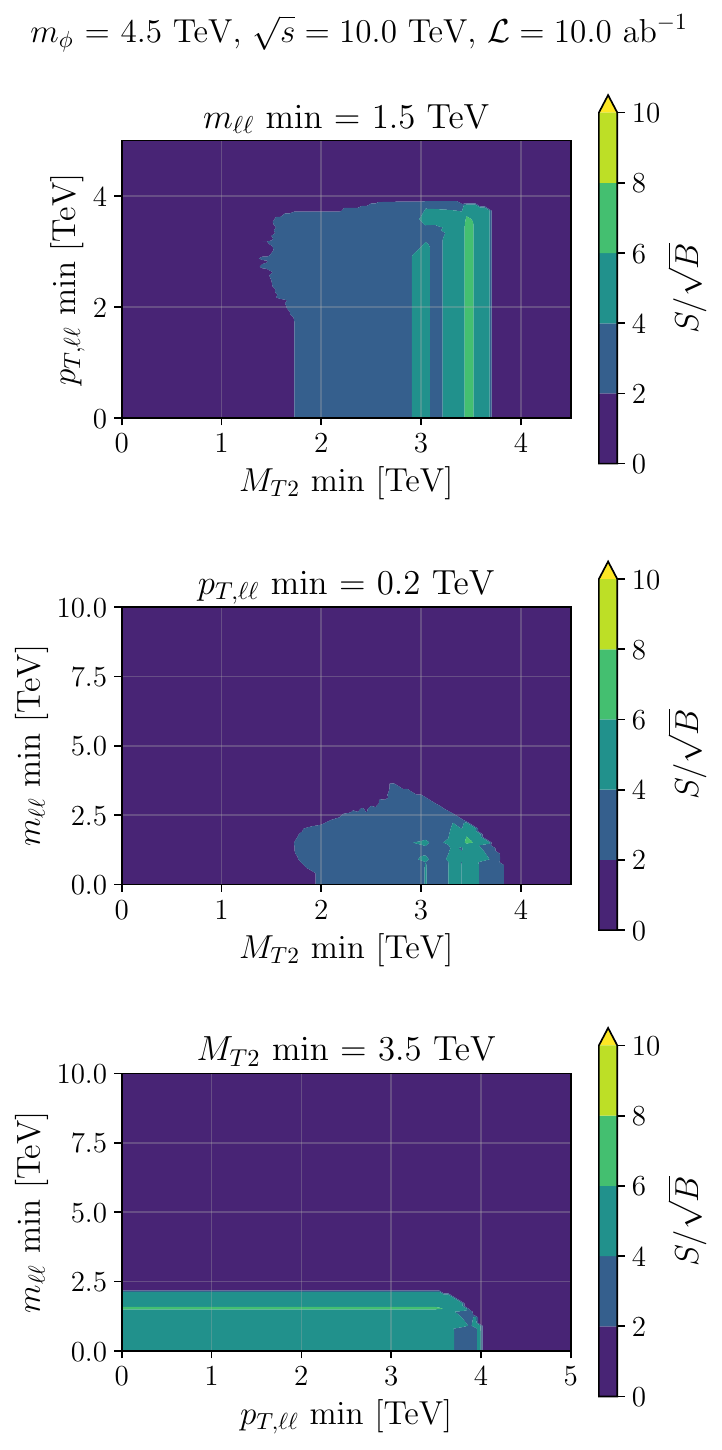}
    }
    \caption{Contours of significance of the signal (signal count $S$ over square root of background $\sqrt{B}$) for different minimum values on each observable. This is the same as Figure~\ref{fig:histograms_2D_prompt_3}, but for $\phi$ masses 1 TeV (\textbf{left}) and 4.5 TeV (\textbf{right}), and at a MuC with center of mass energy of $\sqrt{s}=10$~TeV. We find large significances are attainable. These figures inform the cuts used in definition of the signal regions.}
    \label{fig:histograms_2D_prompt_10}
\end{figure}

To better determine what cuts can optimize the reach of a MuC collider, in Figure~\ref{fig:histograms_2D_prompt_3} (Figure~\ref{fig:histograms_2D_prompt_10}) we show 2D histograms of the ratio of the signal ($S$) to square root of the background ($B$) for a few different $\phi$ masses and for center of mass energy of 3~TeV (10~TeV). These histograms inform us about what lower bounds on each kinematic parameter yields the best reach for any given mediator mass and collider's center of mass energy.

\begin{figure}
    \centering
    \resizebox{1\columnwidth}{!}{
    \includegraphics{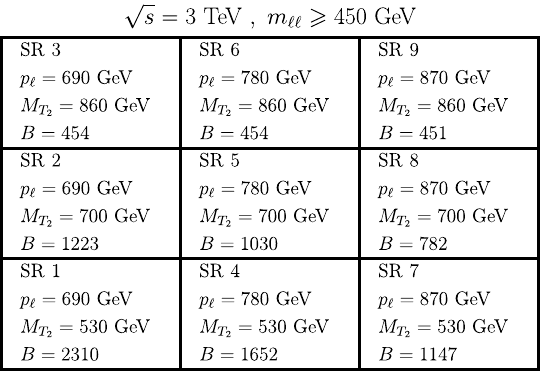}
    \hspace{0.8cm}
    \includegraphics{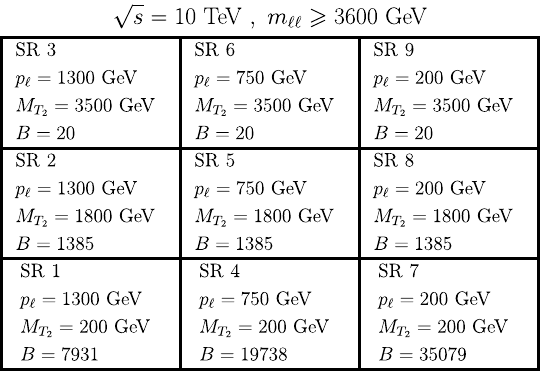}
    }\\
    \vspace{0.3cm}
    \resizebox{1\columnwidth}{!}{
    \includegraphics{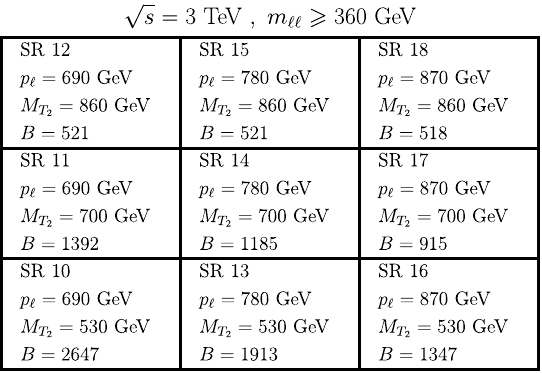}
    \hspace{0.8cm}
    \includegraphics{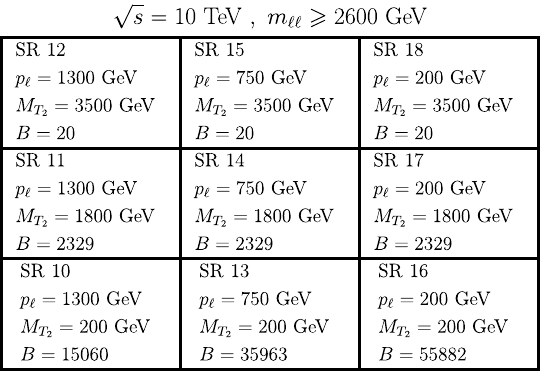}
    }\\
    \vspace{0.3cm}
    \resizebox{1\columnwidth}{!}{
    \includegraphics{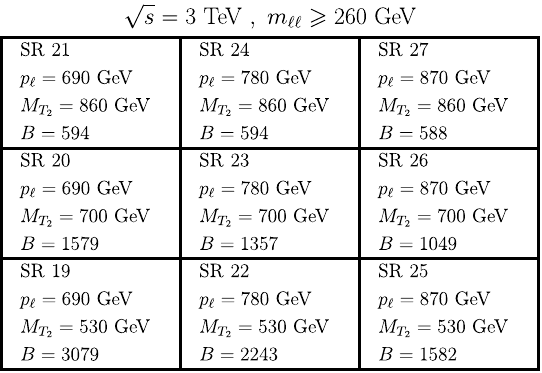}
    \hspace{0.8cm}
    \includegraphics{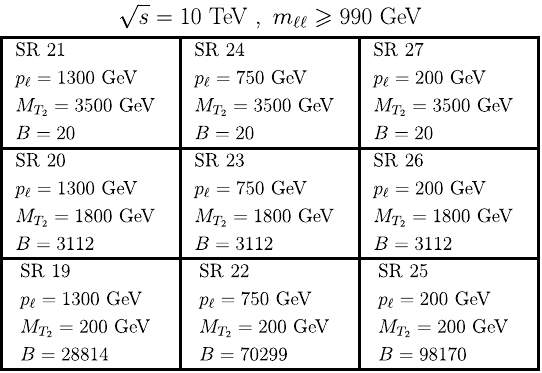}
    }
    \caption{Signal regions (SRs) used in our proposed search for $\sqrt{s}=3$~TeV (\textbf{left}) and for $\sqrt{s}=10$~TeV (\textbf{right}). The cuts used on each kinematic variable and the SM background in each region is reported. Determination of the cuts are informed by the most sensitive cuts on various extreme masses in the parameter space from Eqs.~\eqref{eq:cuts_optimize1_10}--\eqref{eq:cuts_optimize1.4_3}. }
    \label{fig:SRdefs}
\end{figure}

Based on these results, we can define a set of signal regions that together are sensitive to all values of $m_\phi$.
The definition of signal regions we use and their background $B$ is included in Figure~\ref{fig:SRdefs}.

\begin{figure}
    \centering
    \resizebox{0.9\columnwidth}{!}{
    \includegraphics{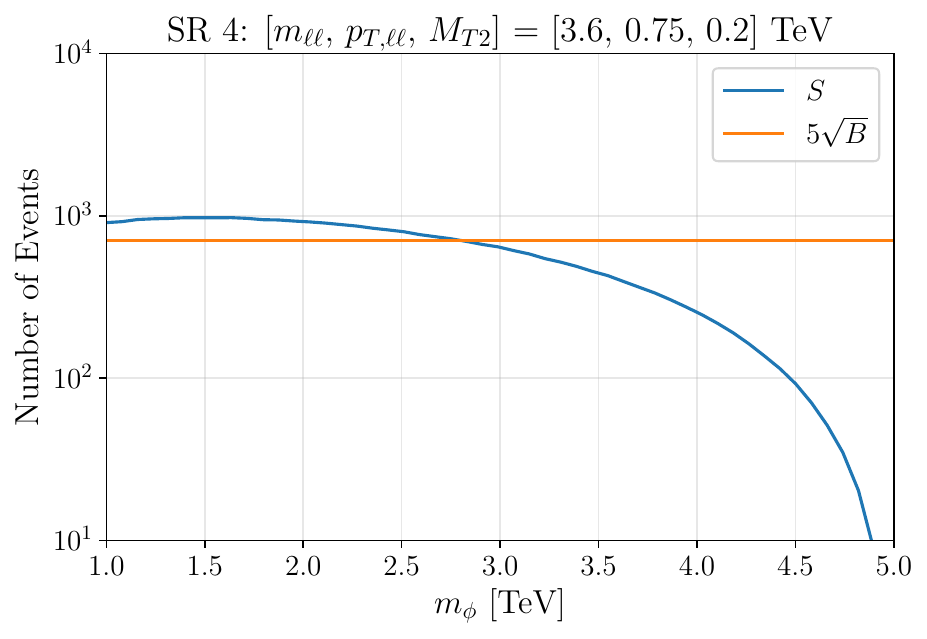}
    \includegraphics{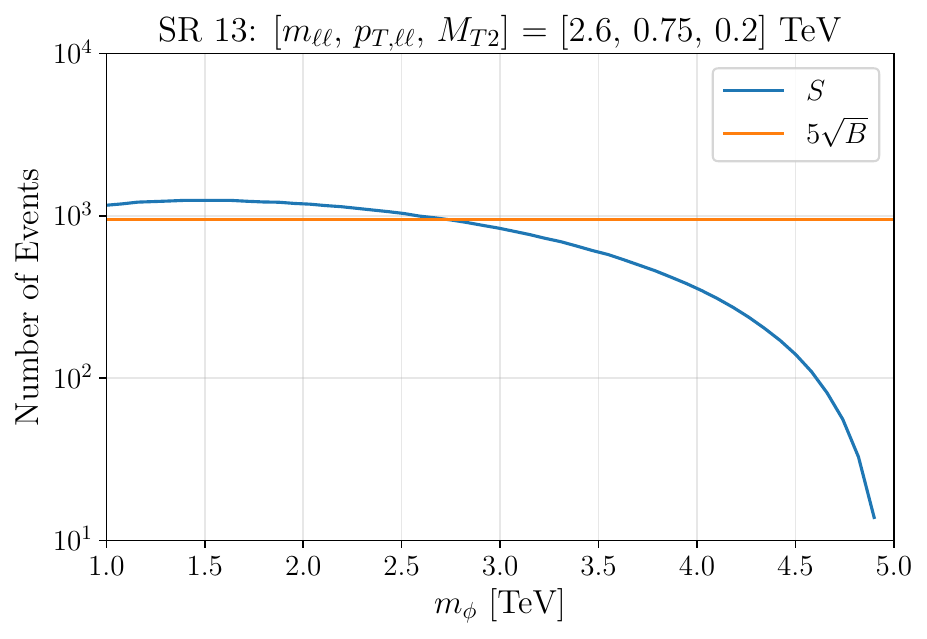}
    }\\
    \resizebox{0.9\columnwidth}{!}{
    \includegraphics{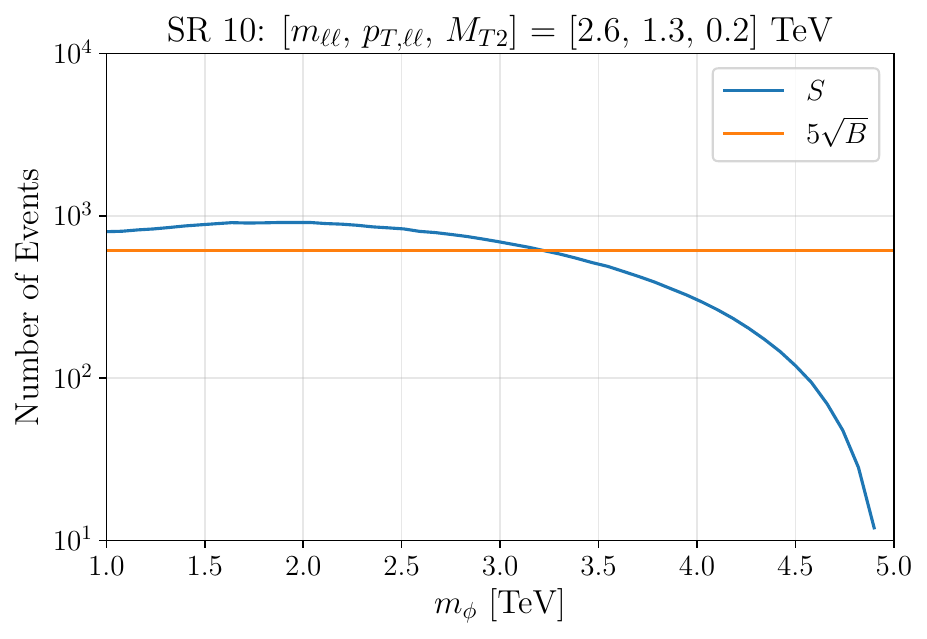}
    \includegraphics{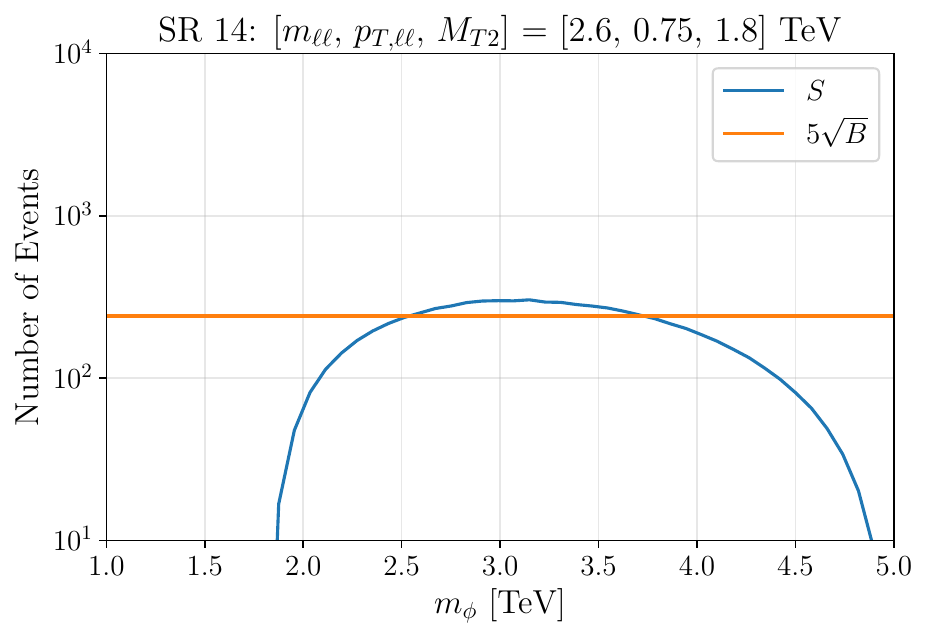}
    }
    \caption{Signal yield and the SM background of four sample signal regions with similar cuts, for different mediator masses. For each mass if a signal region has $S \geqslant 5 \sqrt{B}$ signal events it can give rise to a discovery (neglecting systematics). We find that the $M_{T2}$ cut controls the lowest mediator mass each signal region is sensitive too. Different $m_{\ell\ell}$ cuts affect the total number of events, while leaving the overall dependence on mediator mass unchanged. Comparing the signal regions with similar cuts we find that increasing the $p_{T,\ell\ell}$ cut shifts the signal distribution to higher $\phi$ masses. }
    \label{fig:SRs}
\end{figure}

To further illustrate the effect of each cut as a function of $\phi$ mass, in Figure~\ref{fig:SRs} we show the signal yield and the background for four sample signal regions with very similar cuts. 
These figures indicate that an $M_{T2}$ cut is the lower bound on the mediator masses that populate a signal region, as expected \cite{Lester:1999tx}. 
The $p_{T,\ell\ell}$ and $m_{\ell\ell}$ cuts also reduce the SM and total signal yield in a signal region. Increasing the $p_{T,\ell\ell}$ cut shifts the distribution to higher $m_\phi$ masses as well.

\section{More Details on the LLP Search}
\label{app:LLP}

\begin{figure}
    \centering
    \resizebox{\columnwidth}{!}{
    \includegraphics{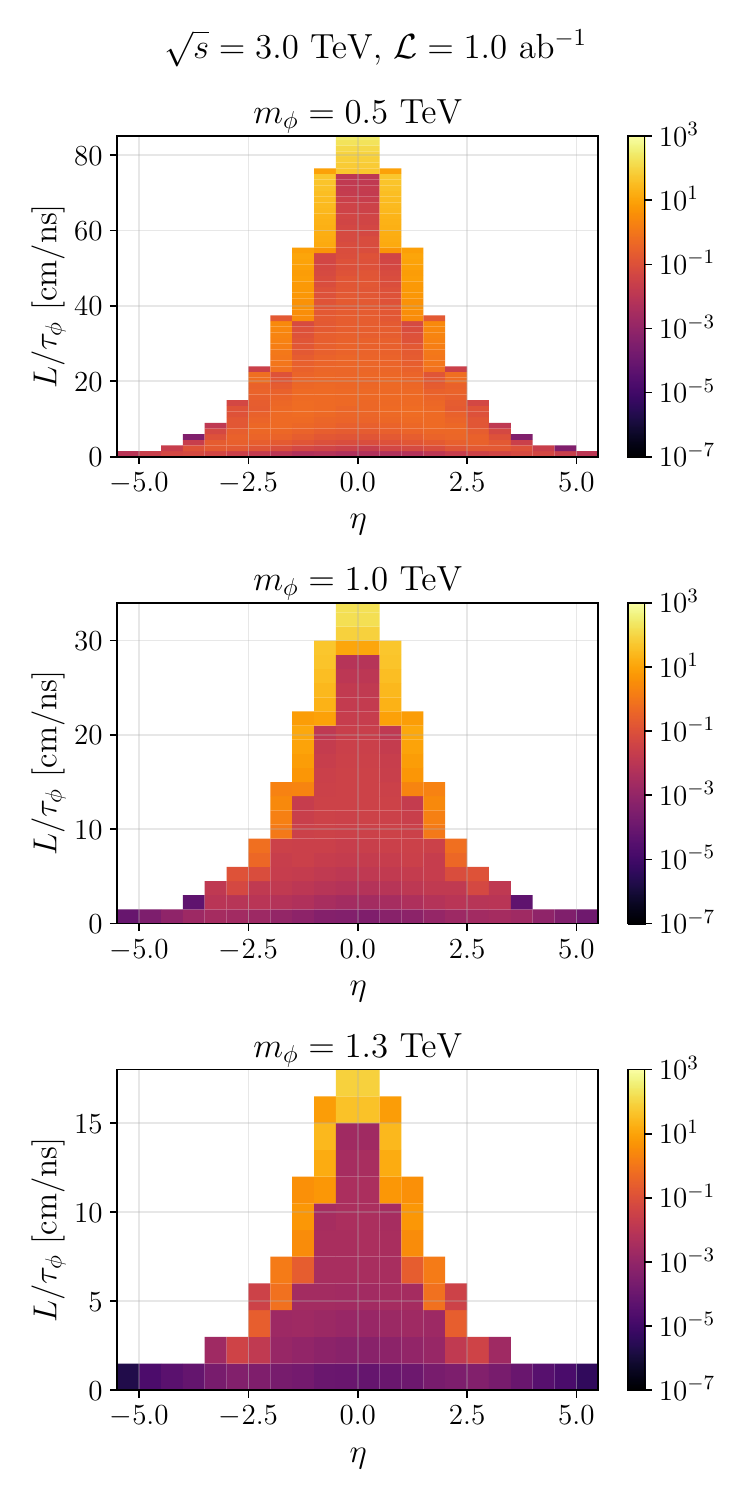}
    \includegraphics{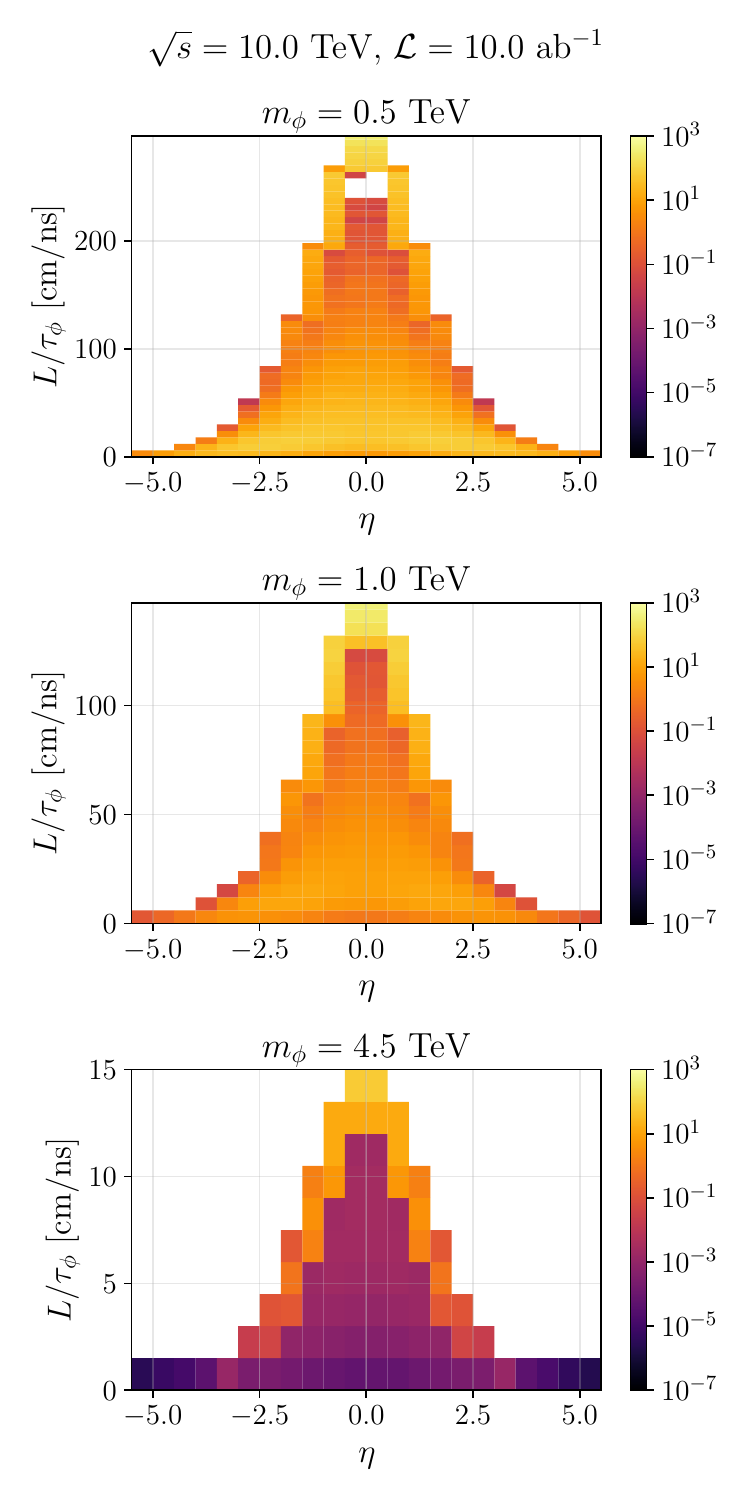}
    }
    \caption{Joint distribution of events in $\eta-L/\tau_\phi$ for three different mediator masses (different rows) at a 3 TeV MuC (\textbf{left}) or a 10 TeV MuC (\textbf{right}). The bin sizes are all 0.5 for $\eta$ and 1.5 cm/ns for $L/\tau_\phi$ (except for $\sqrt{s} = 10$ TeV, $m_\phi = 0.5$ and $1$ TeV where they are 6 cm/ns). Events at the largest $L/\tau_\phi$ values are due to the DY initial channel. The double-peak feature in the distribution (at large and small $L$ values) is manifested as well.   }
    \label{fig:2DLdistseta}
\end{figure}

\begin{figure}
    \centering
    \resizebox{\columnwidth}{!}{
    \includegraphics{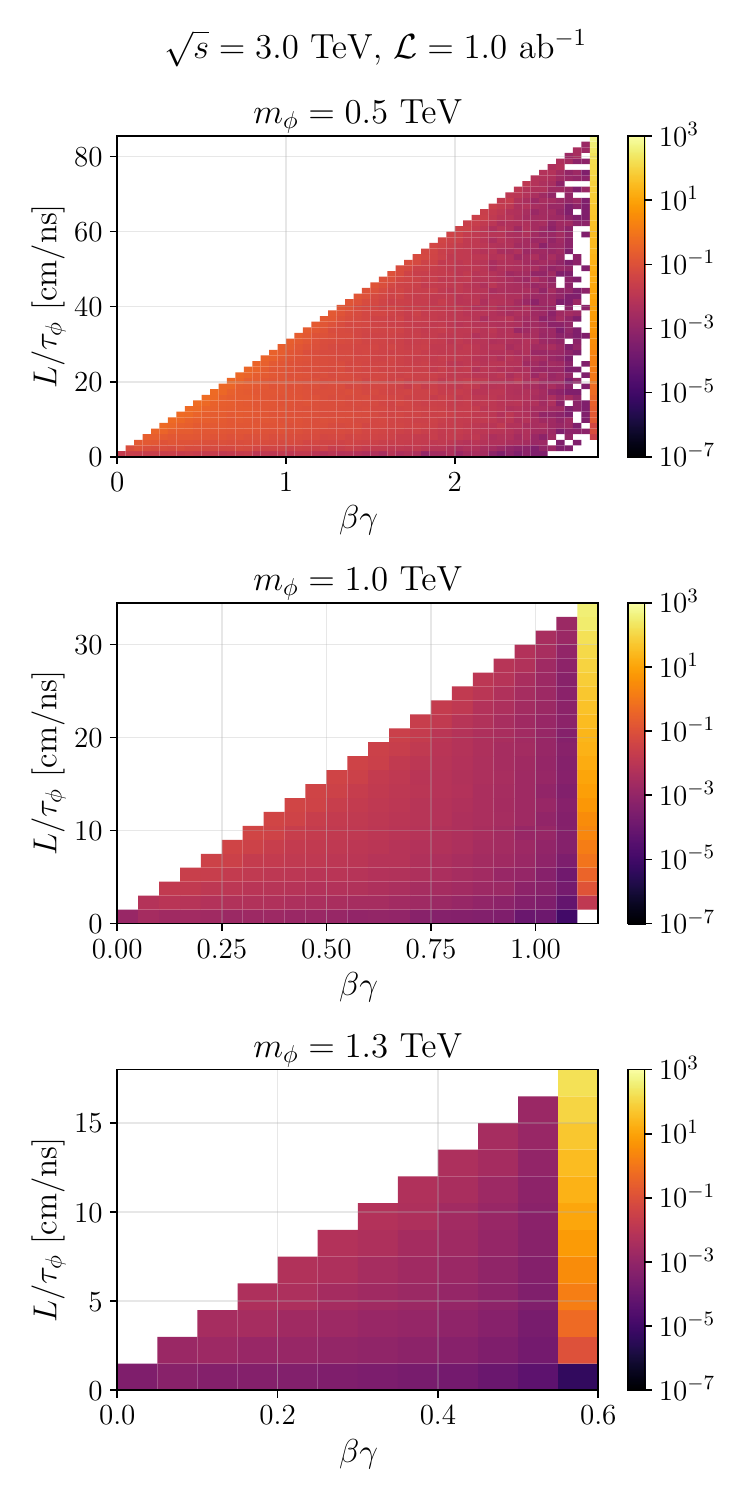}
    \includegraphics{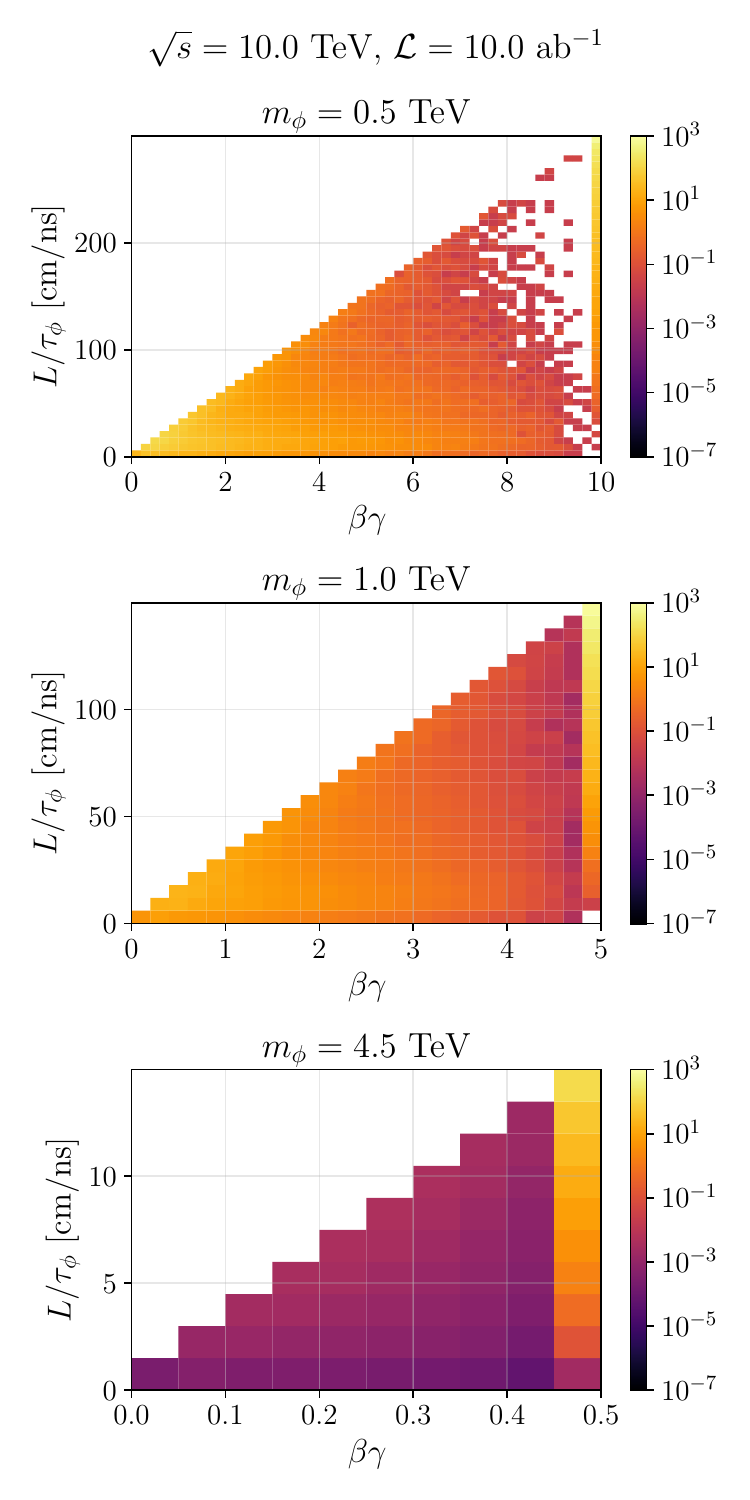}
    }
    \caption{Joint distribution of events in $\beta \gamma-L/\tau_\phi$ for three different mediator masses (different rows) at a 3 TeV MuC (\textbf{left}) or a 10 TeV MuC (\textbf{right}). Bin sizes are ($L/\tau_\phi$, $\beta \gamma$) = (1.5 cm/ns, 0.05) (except $\sqrt{s} = 10$ TeV, $m_\phi = 0.5$ and $1$ TeV where they are ($L/\tau_\phi$, $\beta \gamma$) = (6 cm/ns, 0.2)). Events at the largest $L/\tau_\phi$ values are due to the DY initial channel. The double-peak feature in the distribution (at large and small $L$ values) is manifested as well. }
    \label{fig:2DLdistsbeta}
\end{figure}

For completeness, in this appendix we include the 2D histograms of events in different kinematic observables in the LLP search.
In particular, in Figure~\ref{fig:2DLdistseta} (Figure~\ref{fig:2DLdistsbeta}) we show the 2D histograms of the distribution of events in the transverse distance $L(t=\gamma \tau_\phi)/\tau_\phi$ and their pseudorapidity $\eta$ (their $\beta \gamma$), for a few mass points in the parameter space.
The double-peak feature in Figures~\ref{fig:dist1Dbetaeta}-\ref{fig:dist2Dbetaeta} are still visible here as well. 
The figure also shows that the smaller the mediator mass, the more separated in $L/\tau_\phi$ the two peaks in the distribution.


Figure~\ref{fig:2DLdistseta} also reiterates the concentration of events at small $|\eta|$ values. This allows us to introduce stringent cuts on this variable to cut down the SM background, while preserving majority of signal events. 
The fact that the DY-generated events are more concentrated at smaller $|\eta|$ values is also an underlying cause of the separation of the cluster of events from different channels in Figures~\ref{fig:money_LLP3}--\ref{fig:money_tau_LLP10}.

\begin{figure}
    \centering
    \resizebox{\columnwidth}{!}{
    \includegraphics{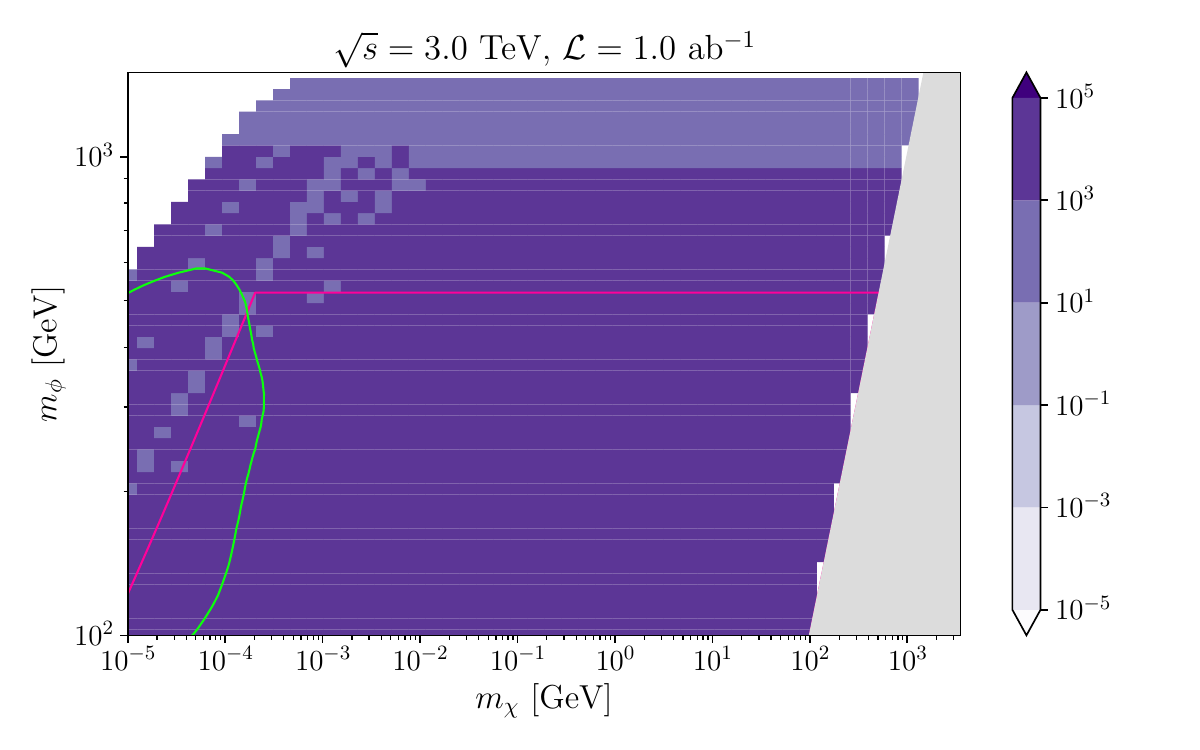}
    \includegraphics{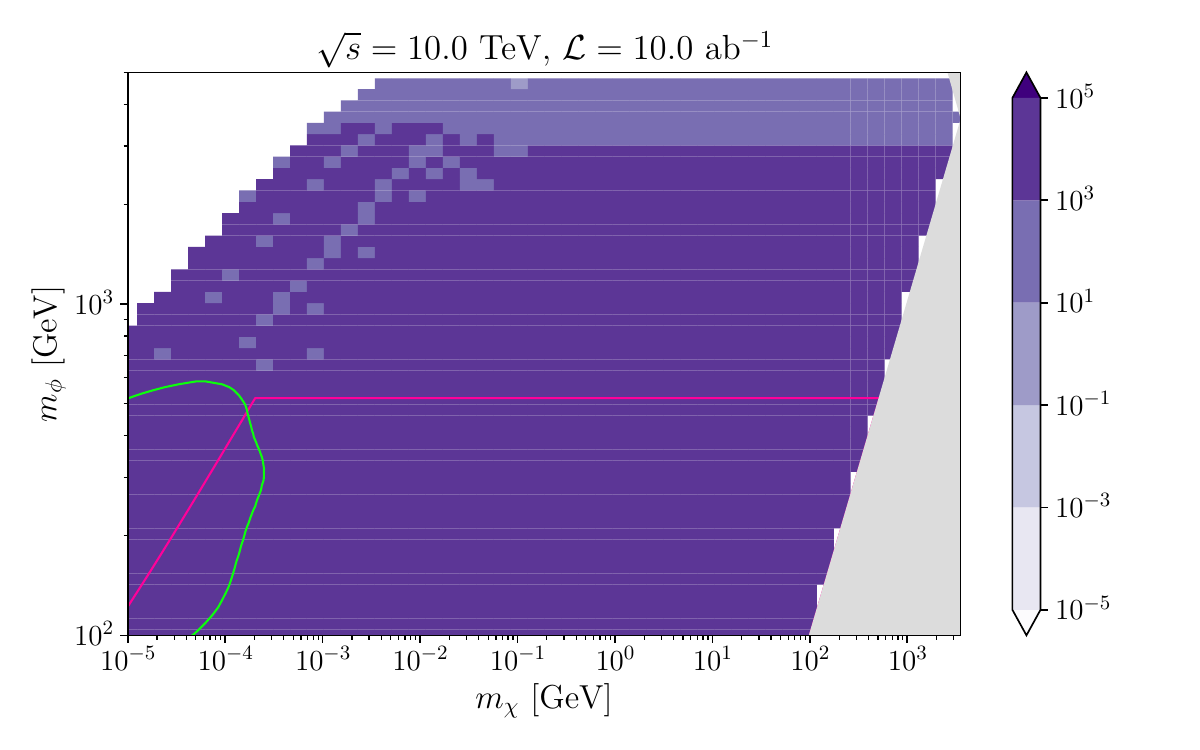}
    }
    \caption{Maximum rate of displaced leptons across different detector regions or stable charged tracks for every point in the parameter space in a 3~TeV (\textbf{left}) and a 10~TeV (\textbf{right}) MuC. The region below the green (pink) line is already ruled out by the LHC search in Ref.~\cite{ATLAS:2020wjh} (Ref.~\cite{CMS:2024qys}).}
    \label{fig:max_LLP}
\end{figure}

\begin{figure}
    \centering
    \resizebox{\columnwidth}{!}{
    \includegraphics{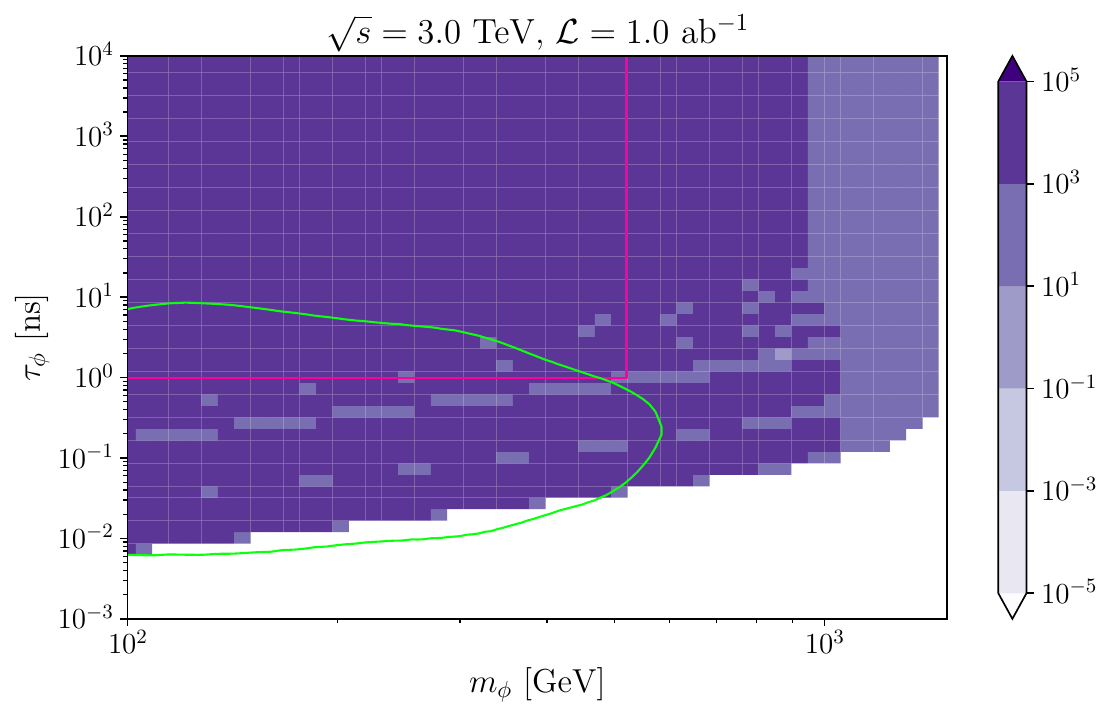}
    \includegraphics{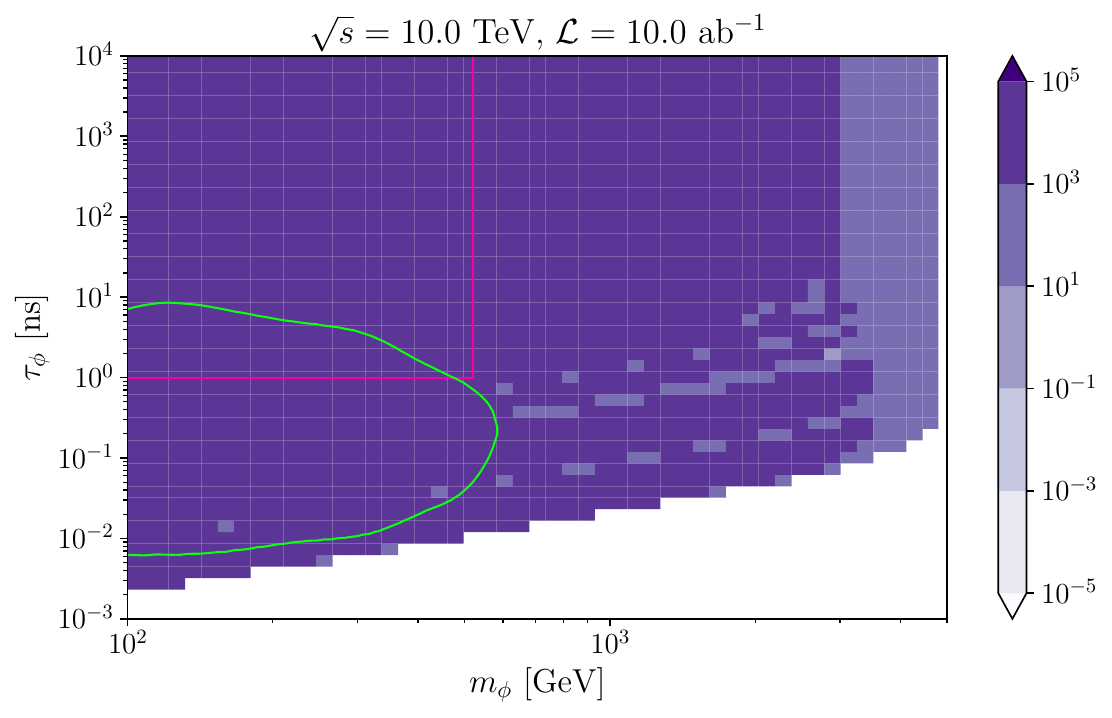}
    }
    \caption{Similar to Figure~\ref{fig:max_LLP} but on the plane of $m_\phi$-$\tau_\phi$ instead. The region below the green (pink) line is already ruled out by the LHC search in Ref.~\cite{ATLAS:2020wjh} (Ref.~\cite{CMS:2024qys}).  }
    \label{fig:max_tau_LLP}
\end{figure}

In Figures~\ref{fig:max_LLP} and \ref{fig:max_tau_LLP} we show the maximum number of decays across different detector segments. 
To make these plots, for each point in the parameter space we simply report the maximum event count from among all panels in Figures~\ref{fig:money_LLP3}--\ref{fig:money_tau_LLP10}. 
We find that for most points in the parameter space we have at least $10^3-10^4$ events in at least one part of the detector (including events that are detector stable and give rise to stable charged tracks).


\end{spacing}

\newpage

\bibliographystyle{utphys}
\bibliography{ref}

\end{document}